\documentclass[final,3p,times]{elsarticle}
\usepackage{hyperref} 
\usepackage{graphicx}
\usepackage{graphbox}
\usepackage{subfigure}
\usepackage{color}
\usepackage{setspace}
\usepackage{subfigure}
\usepackage{multirow}
\usepackage{float}

\usepackage{multirow}                
\usepackage{bm}
\usepackage{amsmath}
\usepackage{amssymb}
\usepackage{makecell}
\usepackage{makecell}
\newcommand{\tabincell}[2]{\begin{tabular}{@{}#1@{}}#2\end{tabular}}

\newcommand{\PreserveBackslash}[1]{\let\temp=\\#1\let\\=\temp}
\newcolumntype{C}[1]{>{\PreserveBackslash\centering}p{#1}}
\newcolumntype{R}[1]{>{\PreserveBackslash\raggedleft}p{#1}}
\newcolumntype{L}[1]{>{\PreserveBackslash\raggedright}p{#1}}

\def\mP{\mathcal{P}}

\journal{Medical Image Analysis}

\biboptions{authoryear}

\begin{document}

\begin{frontmatter}

\title{\textcolor{black}{InDuDoNet+: A Deep
Unfolding Dual Domain Network}\\ \textcolor{black}{for Metal Artifact Reduction in CT Images}}





  \author[1]{Hong Wang}
  \author[1]{Yuexiang Li}
  \author[2]{Haimiao Zhang}
  \author[3,4,5]{Deyu Meng\corref{cor1}}
 \cortext[cor1]{Corresponding author}
  \ead{dymeng@mail.xjtu.edu.cn}
  \author[1]{Yefeng Zheng}
  \author[]{\emph{Fellow, IEEE}}

  \address[1]{Tencent Jarvis Lab, Shenzhen, China}
  \address[2]{Beijing Information Science and Technology University, Beijing, China}
\address[3]{Xi'an Jiaotong University, Xi'an, China}
\address[4]{Peng Cheng Laboratory, Shenzhen, China}
\address[5]{Macau University of Science and Technology, Taipa, Macao}

  \begin{abstract}  During the computed tomography (CT) imaging process, metallic implants within patients often cause harmful artifacts, which adversely degrade the visual quality of reconstructed CT images and negatively affect the subsequent clinical diagnosis. For the metal artifact reduction (MAR) task, current deep learning based methods have achieved promising performance. However, most of them share two main common limitations: 1) the CT physical imaging geometry constraint is not comprehensively incorporated into deep network structures; 2) the entire framework has weak interpretability for the specific MAR task; hence, the role of each network module is difficult to be evaluated. To alleviate these issues, in the paper, we construct a novel {\textcolor{black}{deep unfolding}} dual domain network, termed InDuDoNet+,  into which CT imaging process is finely embedded. Concretely, we derive a joint spatial and Radon domain reconstruction model and propose an optimization algorithm with only simple operators for solving it. By unfolding the iterative steps involved in the proposed algorithm into the corresponding network modules, we easily build the InDuDoNet+ with clear interpretability. Furthermore, we analyze the CT values among different tissues, and merge the prior observations into a prior network for our InDuDoNet+, which significantly improve its generalization performance.
  Comprehensive experiments on synthesized data and clinical data substantiate the superiority of the proposed methods as well as the superior generalization performance beyond the current state-of-the-art (SOTA) MAR methods. Code is available at \url{https://github.com/hongwang01/InDuDoNet_plus}.
  \end{abstract}

 
\end{frontmatter}




\section{Introduction}
Computed tomography (CT) images reconstructed from X-ray projections have been extensively adopted in clinical diagnosis and treatment planning. Unfortunately, metallic implants within patients always lead to the missing projection data and the captured CT images present streaky and shading artifacts, which negatively affect the clinical diagnosis~\citep{de1999metal,park2018ct}. Hence, it is worthwhile to develop effective metal artifact reduction (MAR) methods for CT image reconstruction. 

In recent years, many traditional methods~\citep{mehranian2013x, chang2018prior,jin2015model,lemmens2008suppression,kalender1987reduction,meyer2010normalized,wang2013metal} have been proposed for the MAR task, which can be mainly divided into three categories, \emph{i.e.,} iterative reconstruction, sinogram domain MAR, and image domain MAR. Particularly, iterative algorithms aim at designing some hand-crafted regularizers, such as total variation~\citep{schiffer2014sinogram,zhang2016iterative} and sparsity constraints in the wavelet domain~\citep{zhang2018reweighted}, and formulating them into the algorithm optimization to constrain the solution space. Due to the subjective prior assumptions, these approaches cannot finely represent complicated and diverse metal artifacts in clinical applications. The sinogram domain based methods regard metal-affected regions (\emph{i.e.}, metal trace in sinogram) as missing data and fill them via linear interpolation (LI)~\citep{kalender1987reduction} or forward projection (FP) of a prior image~\citep{meyer2010normalized,wang2013metal}. Yet, these surrogate data in the metal trace often do not properly meet the CT imaging geometry constraint, which causes secondary artifacts tangent to the metallic implants in the reconstructed CT images. The image domain based methods directly utilize some image processing technologies to overcome the adverse artifacts, which often have some limitations for the MAR performance improvement~\citep{karimi2015metal,soltanian1996ct}.

Driven by the tremendous success of deep learning (DL) in medical image reconstruction and analysis~\citep{ronneberger2015u,wang2018image}, researchers began to apply the convolutional neural network (CNN) for MAR~\citep{wang2021dicdnet,zhang2018convolutional,lin2019dudonet,liao2019adn,yu2020deep,lyu2020dudonet++}. The existing DL-based MAR methods can be roughly grouped into three research lines, \emph{i.e.,} sinogram domain enhancement, image domain enhancement, and dual domain (joint sinogram and CT image) enhancement. Specifically, the sinogram-domain-based approaches utilize deep learning networks to directly recover metal-affected sinogram data~\citep{park2018ct,ghani2019fast,liao2019generative} or adopt the forward projection (FP) of a prior image recovered by CNNs to refine the sinogram~\citep{gjesteby2017deep, zhang2018convolutional}. {\textcolor{black}{After obtaining the restored sinogram, the artifact-reduced CT image can be got by executing the filtered back-projection (FBP) operation. }}The image-domain-based approaches exploit deep CNNs to directly learn the mapping function from metal-corrupted CT images to the clean ones based on residual learning~\citep{huang2018metal} or adversarial learning~\citep{wang2018conditional,liao2019adn}. {\textcolor{black}{For the research line of dual domain, recent studies~\citep{lin2019dudonet,yu2020deep,lyu2020dudonet++, zhou2022dudodr} generally adopt two U-shape enhancement sub-networks for accomplishing the reconstruction of sinogram domain and CT image domain, respectively. Every sub-network in most of deep MAR methods is heuristically constructed based on the off-the-shelf network blocks and it can be regarded as a general image enhancement module, which can also be applied to other image restoration tasks. For the data transformation from one domain to another domain between the two sub-networks, differentiated FP layer and filtered back-projection layer are adopted. Due to the joint utilization of sinogram and CT image, the dual-domain-based research line can generally obtain better MAR performance.}}


Attributed to the robust feature representations learned by CNN, the DL-based MAR techniques generally outperform the conventional methods based on hand-crafted features. However, the existing DL-based MAR techniques still share some limitations: 1) most of them regard MAR as the general image restoration problem, which put less emphasis on the full embedding of the inherent physical geometry constraints across the entire learning process. {\textcolor{black}{For example, for the dual-domain-based deep MAR methods, most of them adopt two enhancement networks to separately repair the sinogram and CT image and do not fully model the physical degradation process underlying the specific MAR task, which is more like a data-driven methodology.}} The physical imaging constraints, however, should be potentially helpful to further boost the performance of MAR; 2) most of the existing approaches rely on the off-the-shelf DL toolkits to build different network architectures, which lack sufficient model interpretability for the specific MAR task. {\textcolor{black}{For example, for the work~\citep{yu2020deep}, the sinogram-domain network directly executes the concatenation between the sinogram data and metal trace and then take it as the network input. Based on such heuristic network design manner, the intrinsic role of every network module for MAR is relatively difficult to be explicitly analyzed.}} {\textcolor{black}{Against the aforementioned issues, we model the inherent physical degradation process underlying the specific MAR task}} and propose a novel dual domain network framework, termed InDuDoNet+, for the MAR task. The proposed framework sufficiently embeds the intrinsic imaging geometry model constraints into the process of mutual learning between spatial (image) and Radon (sinogram) domains, which is flexibly integrated with the dual-domain-related prior learning. {\textcolor{black}{Our method is expected to possess advantages of both model-driven and data-driven methodologies where the iterative learning between the sinogram domain and CT image domain gradually and alternately proceeds with clear interpretability and the physical imaging mechanism passes through the entire network structure.}} Specifically, our contributions can be mainly summarized as:

\begin{itemize}
    \item For the MAR task, we specifically propose a concise dual domain reconstruction model and utilize the proximal gradient technique~\citep{beck2009fast} to design an optimization algorithm. Different from traditional solvers~\citep{zhang2018reweighted} containing heavy operations (\emph{e.g.,} matrix inversion), the proposed algorithm is composed of only simple computations (\emph{e.g.,} point-wise multiplication), largely facilitating its easy deep unfolding to a network architecture.
    \item {\textcolor{black}{By unfolding the iterative algorithm, we easily construct a dual domain network, called InDuDoNet+. The specificity of InDuDoNet+ lies in the corresponding relationship between its neural network modules and algorithm operations, resulting in a clear working mechanism.}}
    \item To further improve the generalization performance, we embed the prior characteristics of metal-corrupted CT images into an elaborately designed Prior-net involved in InDuDoNet+. Besides, the network capacity is largely shrunk by a simple weight net, which finely benefits the computational efficiency and generalization ability.
    \item Comprehensive experiments are executed on synthetic and clinical data and they fully substantiate the effectiveness of our method as well as its superior generalization ability beyond the current state-of-the-art MAR methods. Besides, more analysis and verifications are given, which show the good potential of our methods for real applications.
\end{itemize}

An early version~\citep{wang2021indudonet} of this work was presented at a conference. This paper extends the previous work substantially with following improvements: 1) In previously proposed InDuDoNet~\citep{wang2021indudonet}, {\textcolor{black}{the Prior-net (refer to Fig.~\ref{fignet}) is almost a black-box without fully considering prior knowledge. In contrast, in this work, the designed Prior-net (see Fig.~\ref{figwnet}) is finely integrated with empirical prior observations with clearer working mechanism;}} 2) Compared with the previous InDuDoNet, the network parameters of our InDuDoNet+ are largely shrunk by 
a simple WNet (see Table~\ref{tabtime}), which accordingly boosts its generalization performance (see Sec.~\ref{sec:genera1} and Sec.~\ref{sec:genera2});
3) Apart from the datasets adopted by~\cite{wang2021indudonet}, we further validate the effectiveness of the proposed InDuDoNet+ on two additional datasets. Moreover, we conduct more comprehensive experiments on model verification and module analysis in Sec.~\ref{sec:analysis}, and provide the detailed information on network implementation in Sec.~\ref{sec:impdetail}.

The paper is organized as follows. Sec.~\ref{model} provides the dual domain reconstruction model and the corresponding optimization algorithm.  Sec. \ref{sec:details} constructs the interpretable dual domain network where every network module has specific physical meanings, corresponding to the iterative step involved in the proposed optimization algorithm.  Sec.~\ref{sec:analysis} analyzes the role of Prior-net and designs a model-driven Prior-net which is integrated with the prior knowledge of metal-corrupted CT images.  Sec.~\ref{sec:impdetail} describes the network details. Sec.~\ref{sec:exp} demonstrates the experimental evaluations to validate the superiority of the proposed network. The paper is finally concluded in Sec.~\ref{sec:conclu}.

\section{Joint Spatial and Radon Domain Reconstruction Model}\label{model}
In this section, we first derive the joint spatial and Radon domain reconstruction model for the metal artifact reduction (MAR) task and then propose the corresponding optimization algorithm.

\subsection{Dual Domain Model Formulation}\label{sec:dualmodel}
For an observed metal-affected sinogram {{$Y\in \mathbb{R}^{N_{b}\times N_{p}}$}} with $N_{b}$ and $N_{p}$ as the numbers of detector bins and projection views, respectively, conventional iterative optimization based MAR methods are generally formulated as:
\begin{equation}\label{o1}
    \min_{{X}}\left\|(\bm{1}-Tr)\odot (\mP X-Y)\right\|_{F}^{2} + \lambda g(X),
\end{equation}
where {{$X\in \mathbb{R}^{H\times W}$}} with $H$ and $W$ as the height and width of the CT image $X$, respectively, is the expected metal-free clean CT image (\emph{i.e.,} spatial domain); {\textcolor{black}{$\mP$ represents the forward projection process. In experiments, what we adopt is the fan-beam CT geometry (see Sec.~\ref{sec:exp})}}; $Tr\in \mathbb{R}^{N_{b}\times N_{p}}$ is the binary metal trace (\emph{i.e.}, the metal-corrupted region in sinogram domain);  $\bm{1}\in \mathbb{R}^{N_{b}\times N_{p}}$ is a matrix with all elements as 1; $\odot$ denotes the point-wise multiplication; $g(\cdot)$ is the regularization term for delivering the prior knowledge about $X$; $\lambda$ is a trade-off parameter.

For the spatial and Radon domain mutual learning, we further execute the joint regularization on the dual domain and transform problem (\ref{o1}) to:
\begin{equation}\label{o2}
    \min_{{S,X}}\left\|\mP X-S\right\|_{F}^{2} +\alpha \left\|(\bm{1}-Tr)\odot (S-Y)\right\|_{F}^{2}+ \lambda_{1} g_{1}(S) +\lambda_{2} g_{2}(X),
\end{equation}
where $S\in \mathbb{R}^{N_{b}\times N_{p}}$ is the clean metal-free sinogram (\emph{i.e.,} Radon domain); $\alpha$ is a weight factor to balance the data consistency between spatial domain and Radon domain; $g_{1}(\cdot)$ and $g_{2}(\cdot)$ are regularization functions which represent the prior information of the to-be-estimated $S$ and $X$, respectively; $\lambda_{1}$ and $\lambda_{2}$ are both trade-off parameters. 

{\textcolor{black}{As shown in~\citep{meyer2010normalized,zhang2018reweighted}, it is easier to execute the sinogram completion in a more homogeneous region. Motivated by this, we propose to first normalize the original sinogram via dividing it by a normalization coefficient and then correct the normalized sinogram which has more homogeneous profile than the original sinogram. Correspondingly, we rewrite the to-be-estimated sinogram $S$ as:
\begin{equation}\label{YS}
    S=\widetilde{Y}\odot\widetilde{S},
\end{equation}
where $\widetilde{Y}\in \mathbb{R}^{N_{b}\times N_{p}}$ is the normalization coefficient and $\widetilde{S}\in \mathbb{R}^{N_{b}\times N_{p}}$ is the normalized sinogram.}} {\textcolor{black}{$\widetilde{Y}$ is usually set as the forward projection (FP) of a prior image $\widetilde{X}\in \mathbb{R}^{H\times W}$, \emph{i.e.,} $\widetilde{Y} = \mP \widetilde{X}$~\citep{meyer2010normalized}; In this work, we design a simple model-driven CNN integrated with prior knowledge, called Prior-net, to flexibly learn $\widetilde{X}$ from training data (see Fig.~\ref{figwnet} for more details). Note that from Eq.~\eqref{YS}, we can equivalently derive that  ${S}/{\widetilde{Y}}=\widetilde{S}~~(\widetilde{Y}\neq 0)$. Considering that the division operation is not as stable as the multiplication computation, we thus adopt the form $S=\widetilde{Y}\odot\widetilde{S}$ rather than ${S}/{\widetilde{Y}}=\widetilde{S}$.}} By substituting Eq.~(\ref{YS}) into Eq.~(\ref{o2}), we can derive the final dual domain reconstruction problem as:
\begin{equation}\label{o3}
\min_{{\widetilde{S}, X}}\left\|\mP X-\widetilde{Y}\odot\widetilde{S}\right\|_{F}^{2} +\alpha \left\|(\bm{1}-Tr) \odot  (\widetilde{Y}\odot\widetilde{S} - Y)\right\|_{F}^{2}+ \lambda_{1} g_{1}(\widetilde{S})+\lambda_{2} g_{2}(X).
\end{equation}

From Eq.~(\ref{o3}), it is clear that our goal is to estimate the unknown  $\widetilde{S}$ and $X$ from the observed $Y$. In traditional optimization-based MAR methods, to constrain the solution space, researchers manually designed the regularizers $g_{1}(\cdot)$ and $g_{2}(\cdot)$ and formulated them as explicit forms~\citep{zhang2018reweighted}. However, the pre-specified prior forms cannot always cover the complicated structures of CT images collected from real scenarios. Considering that CNN has powerful representation ability to flexibly fit the prior knowledge, we propose to adopt deep network modules to automatically learn the dual-domain-related prior information $g_{1}(\cdot)$ and $g_{2}(\cdot)$ from training data in a purely end-to-end manner. This strategy has been comprehensively validated to be effective in diverse vision tasks, such as spectral image fusion~\citep{xie2020mhf}, dehazing~\citep{yang2018proximal}, and deraining~\citep{wang2020model, wang2021rcdnet}. The details are described in Sec.~\ref{sec:details}.

\subsection{Optimization Algorithm}\label{sec:solver}
Our goal is to build a deep unfolding network where every network module is possibly corresponding to the iterative steps involved in an optimization algorithm so that the learning process of the entire network is interpretable and controllable. Therefore, it is necessary to design an optimization algorithm for solving problem~(\ref{o3}) efficiently, which contains possibly simple operators that can be easily unfolded into network modules. For the dual domain reconstruction problem~(\ref{o3}), traditional solvers~\citep{zhang2018reweighted} are always composed of complicated computations, such as matrix inversion, which makes it challenging to accomplish the transformation from iterative processes to network units. Hence, we prefer to design a new optimization algorithm only containing simple iterative computations for the problem~(\ref{o3}). Specifically, we rely on a proximal gradient technique~\citep{beck2009fast} to alternately update $\widetilde{S}$ and $X$.  The details are given in the following:


\vspace{3mm}
\noindent\textbf{Updating $\widetilde{S}$}: 
At the $n$-th iteration, the normalized sinogram $\widetilde{S}$ can be updated by solving the quadratic approximation~\citep{beck2009fast} of problem (\ref{o3}) about $\widetilde{S}$, expressed as:
\begin{equation}\label{minqs}
  \min_{\widetilde{S}} \frac{1}{2} \left\| \widetilde{S} -\left( \widetilde{S}_{n-1}- \eta_{1}\nabla f\left(\widetilde{S}_{n-1}\right) \right) \right\|_F^2 + \lambda_{1}\eta_{1} g_{1}(\widetilde{S}),
\end{equation}
where {{$\widetilde{S}_{n-1}$}} is the updated result after $(n-1)$ iterations; $\eta_{1}$ is the stepsize parameter; {{$f\left(\widetilde{S}_{n-1}\right) =  \left\|\mP X_{n-1} - \widetilde{Y}\odot\widetilde{S}_{n-1}\right\|_{F}^{2} +\alpha \left\|(1-Tr)\odot(\widetilde{Y}\odot\widetilde{S}_{n-1} - Y)\right\|_{F}^{2}$}}.
For general regularization terms \citep{donoho1995noising}, the solution of Eq. (\ref{minqs}) can be derived as:
\begin{equation}\label{sols}
 \widetilde{S}_{n} = \mbox{prox}_{\lambda_{1}\eta_{1}}\left(\widetilde{S}_{n-1}  -  \eta_{1}\nabla f\left(\widetilde{S}_{n-1}\right)   \right),
\end{equation}
where
\begin{equation}\label{delats}
\nabla f\left(\widetilde{S}_{n-1}\right) = \widetilde{Y}\odot\left(\widetilde{Y}\odot\widetilde{S}_{n-1}-\mP X_{n-1}\right)+\alpha\left(\bm{1}-Tr\right)\odot\widetilde{Y}\odot\left(\widetilde{Y}\odot\widetilde{S}_{n-1} - Y\right).
\end{equation}
By substituting Eq.~(\ref{delats}) into Eq.~(\ref{sols}),
we can easily get the updating rule of $\widetilde{S}$ as:
\begin{equation}\label{rules}
    \begin{split}
        \widetilde{S}_{n} &= \mbox{prox}_{\lambda_{1}\eta_{1}} \left(\widetilde{S}_{n-1}  -  \eta_{1}\left(\widetilde{Y}\odot \left(\widetilde{Y}\odot\widetilde{S}_{n-1} - \mP X_{n-1}\right) + \alpha\left(\bm{1}-Tr\right)\odot\widetilde{Y}\odot \left(\widetilde{Y}\odot\widetilde{S}_{n-1} - Y\right)   \right) \right) \\
         &\triangleq \mbox{prox}_{\lambda_{1}\eta_{1}} \left(\widehat{S}_{n-1}\right),
    \end{split}
\end{equation}
\normalsize
where $\mbox{prox}_{\lambda_{1}\eta_{1}}(\cdot)$ is the proximal operator dependent on the regularization function $g_{1}(\cdot)$. Instead of adopting fixed hand-crafted image priors~\citep{zhang2016iterative,zhang2018reweighted}, we adopt convolutional network modules to automatically learn the implicit $\mbox{prox}_{\lambda_{1}\eta_{1}}(\cdot)$ from training data (detailed in Sec.~{\ref{sec:details}}).

\vspace{3mm}
\noindent\textbf{Updating $X$}: Similarly, the metal-free CT image $X$ can be updated by solving the quadratic approximation of Eq.~(\ref{o3}) with respect to $X$, written as:
\begin{equation}\label{minqx}
  \min_{X}\frac{1}{2} \Big\| X - \left( X_{n-1}- \eta_{2}\nabla h\left(X_{n-1}\right)   \right) \Big\|_F^2 + \lambda_{2}\eta_{2} g_{2}(X),
\end{equation}
\normalsize
where {\normalsize{$\nabla h\left(X_{n-1}\right) = \mP^{T}\left(\mP X_{n-1}-\widetilde{Y}\odot\widetilde{S}_{n}\right)$}}. Thus, the updating formula of $X$ is expressed as:
\begin{equation}\label{rulex}
    X_{n} = \mbox{prox}_{\lambda_{2}\eta_{2}}\left(X_{n-1}- \eta_{2}\mP^{T}\left(\mP X_{n-1}-\widetilde{Y}\odot\widetilde{S}_{n}\right)\right)\triangleq \mbox{prox}_{\lambda_{2}\eta_{2}}\left(\widehat{X}_{n-1}\right),
\end{equation}
\normalsize
where $\mbox{prox}_{\lambda_{2}\eta_{2}}(\cdot)$ is  the proximal operator related to the prior form $g_{2}(\cdot)$ about $X$. 

As seen, the entire iterative optimization algorithm is composed of Eqs.~(\ref{rules}) and (\ref{rulex}). Both alternative updating steps only contain simple operators, making it easy to execute the unfolding process and thus correspondingly construct the deep network framework. The details are presented in the following section.

\begin{figure*}[t]
  \begin{center}
     \includegraphics[width=1\linewidth]{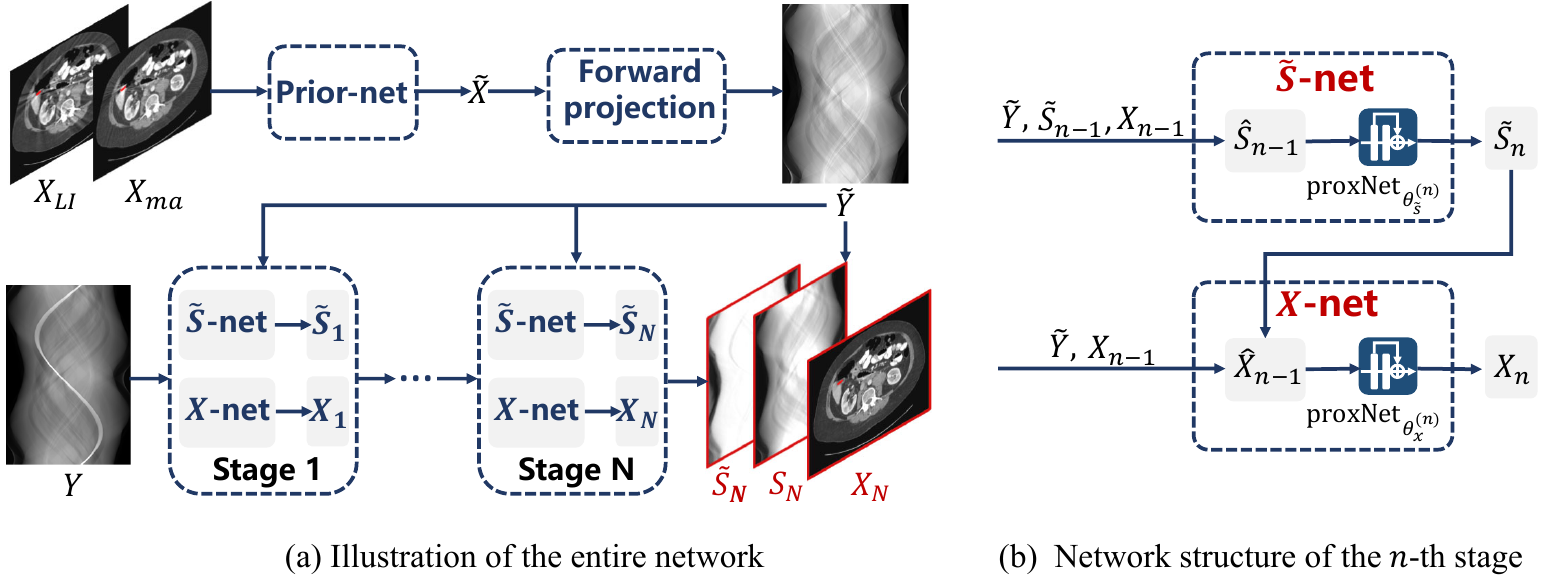}
  \end{center}
  \vspace{-3mm}
     \caption{(a) The proposed network architecture (\emph{i.e.}, InDuDoNet+) consists of a Prior-net, $N$-stage $\widetilde{S}$-net, and $N$-stage $X$-net. It outputs the normalized sinogram $\widetilde{S}_{N}$, sinogram $S_{N}$, and image $X_{N}$. (b) The detailed structure at the $n$-th stage, in which $\widetilde{S}_{n}$ and $X_{n}$ are successively updated by $\widetilde{S}$-net and $X$-net, respectively, based on the algorithm in Eqs.~(\ref{rules}) and (\ref{rulex}).}
  \label{fignet}
\end{figure*}
\section{\textcolor{black}{Deep Unfolding Dual Domain Network}}\label{sec:details}
In many recent studies~\citep{yang2017admm,yang2018proximal, wang2020model,liu2022low, wang2022adaptive, wang2022orientation, fu2022kxnet}, deep unfolding techniques have achieved great success and the fine interpretability of unfolding networks has been substantiated. Motivated by these, in this section, we aim to specifically construct a deep unfolding network, namely InDuDoNet+, for the MAR task.

Specifically, the pipeline of the proposed InDuDoNet+ is illustrated in Fig.~\ref{fignet} (a). As seen, the entire network structure is composed of Prior-net with parameter $\theta_{prior}$ for the prior image $\widetilde{X}$ estimation, $N$-stage $\widetilde{S}$-net with parameter $\theta_{\widetilde{s}}^{(n)}$ for the $\widetilde{S}$ estimation, and $N$-stage $X$-net with parameter $\theta_x^{(n)}$ for the $X$ estimation. At every stage, as illustrated in Fig.~\ref{fignet}(b), $\widetilde{S}$-net and $X$-net are step-by-step constructed based on the updating rules as expressed in Eqs.~(\ref{rules}) and (\ref{rulex}). Clearly, the proposed network framework has specific physical interpretability and it is naturally constructed based on the derived optimization algorithm. All the involved parameters, including $\theta_{prior}$, $\{\theta_{\widetilde{s}}^{(n)},\theta_x^{(n)}\}_{n=1}^{N}$, $\eta_{1}$, $\eta_{2}$, and $\alpha$, can be automatically learned from training data in an end-to-end manner.

\vspace{3mm}
\noindent\textbf{Prior-net.} As shown in Fig.~\ref{fignet} (a), Prior-net is utilized to learn $\widetilde{Y}$ and the network input is composed of metal-affected image $X_{ma}$ and linear interpolation (LI) corrected image $X_{LI}$~\citep{kalender1987reduction}, where $X_{ma}$ and $X_{LI}$ are reconstructed from the original metal-corrupted sinogram $Y$ and the linear interpolated sinogram $Y_{LI}$~\citep{kalender1987reduction}, respectively. The architecture of Prior-net will be discussed in details in next section.


\vspace{3mm}
\noindent{\bf $\widetilde{S}$-net and $X$-net.} 
With the sequential updates of $\widetilde{S}$-net and $X$-net, the framework accomplishes the reconstruction of the artifact-reduced sinogram $\widetilde{S}$ and the CT image $X$, respectively. As shown in Fig.~\ref{fignet}(a), the updating process consists of $N$ stages, which correspond to $N$ iterations of the algorithm for solving problem~(\ref{o3}). Each stage shown in Fig.~\ref{fignet}(b) is correspondingly built by unfolding the iterative rules in Eqs.~(\ref{rules}) and (\ref{rulex}), respectively. Specifically, at the $n$-th stage, $\widehat{S}_{n-1}$ is firstly computed based on Eq.~(\ref{rules}) and then fed to a deep network {\normalsize{$\text{proxNet}_{\theta_{\widetilde{s}}^{(n)}}(\cdot)$}} so as to execute the proximal operator $\mbox{prox}_{\lambda_{1}\eta_{1}}(\cdot)$. Subsequently, we get the updated normalized sinogram as {\normalsize {$\widetilde{S}_{n} = \text{proxNet}_{\theta_{\widetilde{s}}^{(n)}}\left(\widehat{S}_{n-1}\right)$}}. Similarly, for updating the CT image $X$, {\normalsize{$\widehat{X}_{n-1}$}} is firstly computed based on Eq.~(\ref{rulex}) and then fed to a network module {\normalsize{$\text{proxNet}_{\theta_{x}^{(n)}}(\cdot)$}}. Then we obtain the updated artifact-reduced CT image as {\normalsize{$X_{n} = \text{proxNet}_{\theta_{x}^{(n)}}\left(\widehat{X}_{n-1}\right)$}}. Here {\normalsize{$\text{proxNet}_{\theta_{\widetilde{s}}^{(n)}}(\cdot)$}} and {\normalsize{$\text{proxNet}_{\theta_{x}^{(n)}}(\cdot)$}} have the same residual structure, and the details about network implementation are described in Sec.~\ref{sec:impdetail}. 
With the $N$-stage optimization, the proposed InDuDoNet+ can finely recover the normalized sinogram $\widetilde{S}_{N}$, and therefore yield the final sinogram {\normalsize{${S}_{N}$}} by {\normalsize{$\widetilde{Y}\odot\widetilde{S}_{N}$}} (refer to Eq.~(\ref{YS})), and the artifact-reduced CT image {\normalsize{$X_{N}$}}, where $\widetilde{Y}$ is the predicted result of Prior-net.

\vspace{3mm}
{\color{black}{\noindent\emph{\textbf{Interpretability:}}
Similar to conventional optimization methods, every network connection contained in InDuDoNet+ has clear physical meanings and the role of every network module is easily understood. Taking $X$-net as an example, as shown in Fig.~\ref{fignet}(b), the involved computation process is: $X_{n} = \text{proxNet}_{\theta_{x}^{(n)}}\left(X_{n-1}- \eta_{2}\mP^{T}\left(\mP X_{n-1}-\widetilde{Y}\odot\widetilde{S}_{n}\right)\right)$. The corresponding physical meaning is as follows: 
given the previously-estimated CT image $X_{n-1}$, we derive the corresponding sinogram data as $\mathcal{P}X_{n-1}$. With the updating of $\widetilde{S}$-net, we can obtain the estimated sinogram $\widetilde{Y}\odot\widetilde{S}_{n}$ at the current iterative stage. Then the X-net calculates the residual information between the two sinogram data obtained in these two ways as $\left(\mathcal{P}X_{n-1}-\widetilde{Y}\odot\widetilde{S}_{n}\right)$, and extracts the residual information $\mathcal{P}^{T}\left(\mathcal{P}X_{n-1}-\widetilde{Y}\odot\widetilde{S}_{n}\right)$ of CT images with the transposed operation of forward projection matrix $\mathcal{P}$ to update the CT image. Clearly, our network works like a white-box and is expected to possess advantages of both model-driven and data-driven methodologies. Particularly, compared with traditional prior-based methods, our network can more flexibly learn sinogram-related and image-related priors through {\normalsize{$\text{proxNet}_{\theta_{\widetilde{s}}^{(n)}}(\cdot)$}} and {\normalsize{$\text{proxNet}_{\theta_{x}^{(n)}}(\cdot)$}}, respectively, from training data. Compared with deep MAR methods, our framework incorporates both CT imaging constraints and dual-domain-related priors into the network architecture.}}

{\color{black}{
Most of current deep MAR networks are heuristically built based on the off-the-shelf network blocks, such as U-Net, which can be generally applied to a general image restoration task, and the role of every network module is hard to evaluate. However, the proposed InDuDoNet+ is correspondingly constructed under the
guidance of the optimization algorithm with careful data fidelity
term design, and every network module has its own physical meanings, corresponding to specific iterative steps. In this regard, the entire network integrates the interpretability of model-based methods~\citep{zhang2018ista}. Besides,
such interpretability is visually validated by Fig.~\ref{figmodelver}.  Actually, this interpretability is the inherent characteristics of the deep unfolding-based network construction design manner~\citep{zhang2018ista,zhang2020deep}} and it represents that the network design is
integrated with the physical model underlying the task and is equipped with the characteristics
of the traditional model-driven and the currently-popular deep learning-based methodologies.}


\begin{figure*}[t]
  \begin{center}
     \includegraphics[width=1\linewidth]{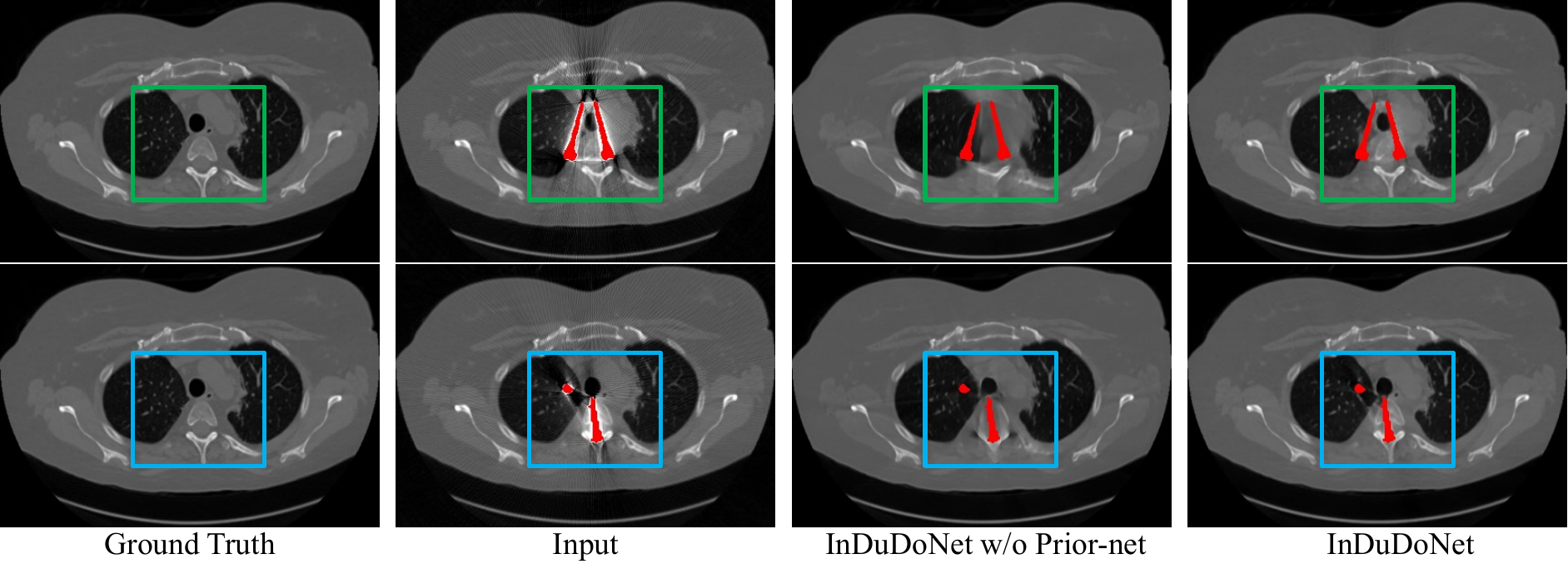}
  \end{center}
  \vspace{-4mm}
     \caption{Performance comparison on the synthesized DeepLesion dataset, where ``InDuDoNet w/o Prior-net'' denotes omitting the Prior-net in InDuDoNet~\citep{wang2021indudonet} and directly learning the sinogram $S$. The red pixels stand for metallic implant.}
  \label{figys}
    \vspace{1mm}
\end{figure*}

\begin{table}[!t]
\centering
\caption{{\textcolor{black}{PSNR (dB) /SSIM of different methods on the synthesized DeepLesion dataset. {\textcolor{black}{The column ``Average'' represents the average PSNR/SSIM on the entire dataset. The column ``STD'' denotes the standard derivation about the PSNR/SSIM results computed on the entire dataset.}}}}}
\setlength{\tabcolsep}{4pt}
\footnotesize
{\textcolor{black}{\begin{tabular}{l|c|c|c|c|c|c|c}
\Xhline{0.6pt}
Methods    & \multicolumn{5}{c|}{Large Metal \quad \quad   \quad\quad  $\longrightarrow$    \quad   \quad\quad \quad         Small Metal}                & Average     & \textcolor{black}{STD} \\
\Xhline{0.6pt}
Input             &24.12/0.6761              &26.13/0.7471              &27.75/0.7659               &28.53/0.7964              &28.78/0.8076              &27.06/0.7586   &\textcolor{black}{2.78/0.0610}           \\
DuDoNet++~\citep{wang2021indudonet}  & 36.17/{0.9784} & {38.34}/{0.9891} & {40.32}/{0.9913}  & 41.56/0.9919 & {42.08}/{0.9921} & {39.69}/{0.9886} &\textcolor{black}{3.86/0.0075}\\
InDuDoNet w/o Prior-net & 33.70/0.9715 &36.56/0.9839 &41.71/0.9929 &44.21/0.9946 &44.73/0.9950 &40.18/0.9876 &\textcolor{black}{5.60/0.0113}\\
InDuDoNet &{36.74}/{0.9801} &{39.32}/{0.9896} &{41.86}/{0.9931} &{44.47}/{0.9942} &{45.01}/{0.9948} &{41.48}/{0.9904} &\textcolor{black}{4.73/0.0088}\\
\Xhline{0.6pt}
\end{tabular}}}
\vspace{-1mm}
\label{tab:claim}
\end{table}

\begin{figure*}[t]
	{{\begin{center}
		{{\includegraphics[width=0.8\linewidth]{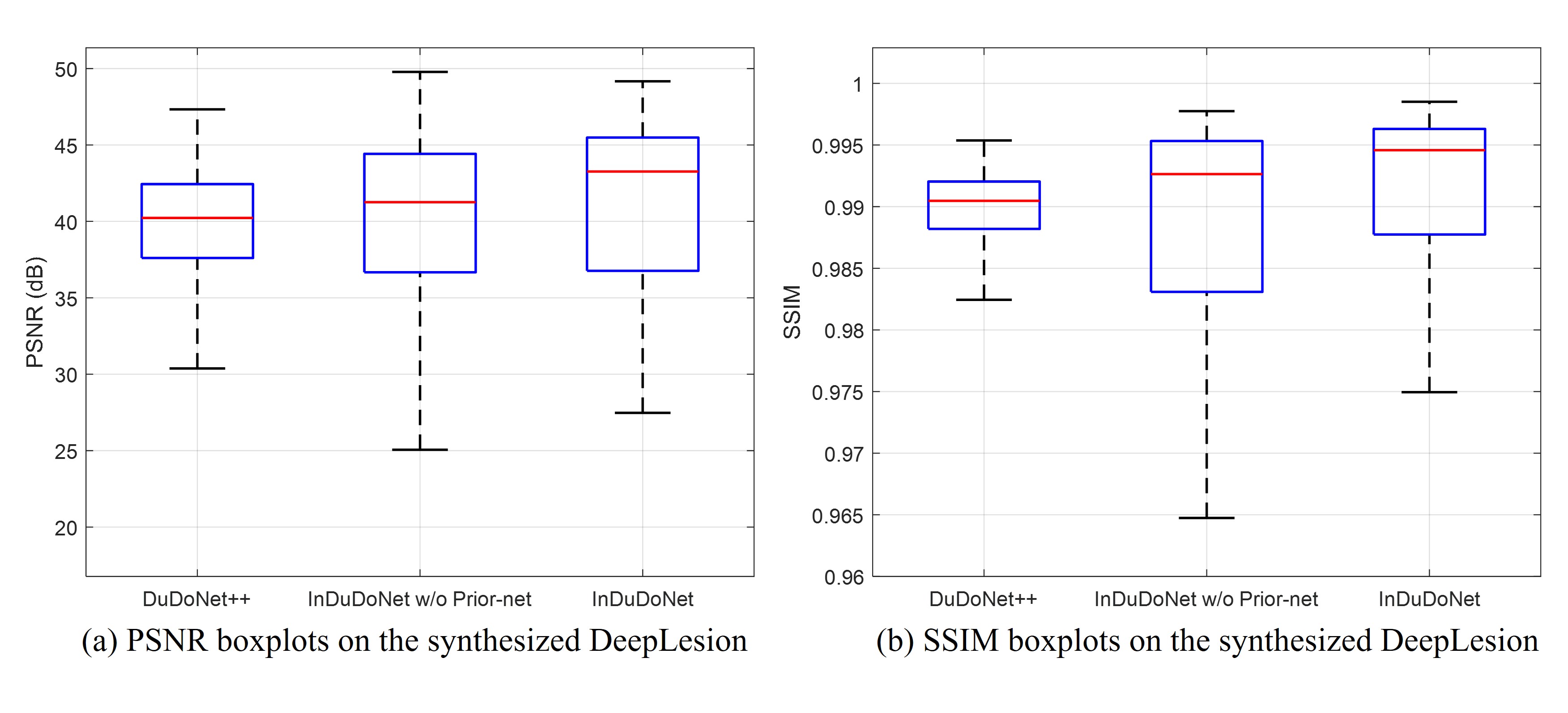}}}
	\end{center}
	\vspace{-8mm}
	\caption{{\textcolor{black}{Boxplots for PSNR/SSIM of different methods in Table~\ref{tab:claim} on the synthesized DeepLesion dataset.}}}
	\label{fig:noysbox}}}
\end{figure*}

\section{Architecture of Prior-net}\label{sec:analysis}
In Sec.~\ref{sec:dualmodel}, we propose to reconstruct the clean sinogram in a normalized manner as given in Eq.~(\ref{o3}) and then correspondingly construct the InDuDoNet+ as shown in Fig.~\ref{fignet} where $\widetilde{S}$-net and $X$-net are both built based on iterative rules. The whole pipeline of InDuDoNet+ is similar to the previous InDuDoNet~\citep{wang2021indudonet}, except the design of Prior-net. Concretely, the Prior-net of InDuDoNet has a similar U-shape architecture~\citep{ronneberger2015u} to the PriorNet in~\citep{yu2020deep} with the depth of four and a halved number of channels. Such a Prior-net is built based on the off-the-shelf U-shape network structure and has weak interpretability.
In this regard, we first comprehensively verify the effectiveness of Prior-net (\emph{i.e.,} the normalization coefficient $\widetilde{Y}$) and further propose a novel architecture for Prior-net with clearer working mechanism and better generalization ability.

\subsection{Analysis on Prior-net}\label{sec:priorver}
To validate the effectiveness of U-shape Prior-net adopted by InDuDoNet~\citep{wang2021indudonet}, we omit $\widetilde{Y}$ and then the corresponding dual domain reconstruction problem is degraded to Eq.~(\ref{o2}). With the same solver (\emph{i.e.}, proximal gradient technique) for Eq.~(\ref{o3}) derived in Sec.~\ref{sec:solver}, we can easily obtain the iterative rules for Eq.~(\ref{o2}) as:
\begin{equation}\label{rules2}
    \begin{split}
       & {S}_{n} = \mbox{prox}_{\lambda_{1}\eta_{1}} \left({S}_{n-1}  -  \eta_{1}\left(\left({S}_{n-1} - \mP X_{n-1}\right) + \alpha\left(\bm{1}-Tr\right)\odot \left({S}_{n-1} - Y\right)   \right) \right)\triangleq \mbox{prox}_{\lambda_{1}\eta_{1}} \left(\widehat{S}_{n-1}\right),\\
       &    X_{n} = \mbox{prox}_{\lambda_{2}\eta_{2}}\left(X_{n-1}- \eta_{2}\mP^{T}\left(\mP X_{n-1}-{S}_{n}\right)\right)\triangleq \mbox{prox}_{\lambda_{2}\eta_{2}}\left(\widehat{X}_{n-1}\right). 
    \end{split}
\end{equation}
\normalsize

By unfolding the iterative process in Eq.~(\ref{rules2}) into network modules, we can easily construct the degraded deep network architecture (\emph{i.e.,} InDuDoNet w/o Prior-net). The difference between InDuDoNet w/o Prior-net and InDuDoNet is that the former has no Prior-net and the sinogram domain network directly updates the sinogram $S$ instead of the normalized sinogram $\widetilde{S}$.
Fig.~\ref{figys} displays the visual comparison between InDuDoNet w/o Prior-net and InDuDoNet on different types of metallic implants, where the clean ground truth CT images are collected from DeepLesion~\citep{yan2018deep} and the metal-corrupted input images are synthesized based on the existing simulation procedure (refer to Sec.~\ref{sec:exp} for details). From the areas marked by green and blue rectangles, it can be easily observed that the artifact-reduced CT images reconstructed by InDuDoNet w/o Prior-net lose lots of detailed information, especially around the metallic implants.
This is mainly attributed to the lack of normalization operation, which makes it challenging to directly recover the non-homogeneous metallic region. In contrast, by using the Prior-net, the normalized profile would be more homogeneous and thus InDuDoNet achieves better reconstruction of details, which validates the effectiveness of normalization operation achieved by Prior-net. {\textcolor{black}{ Table~\ref{tab:claim} lists the average quantitative results and the standard derivations (STDs) of different methods on the entire synthesized DeepLesion dataset and Fig.~\ref{fig:noysbox} presents the corresponding boxplots for PSNR/SSIM.}} From these results, we can easily observe that even without the introduction of Prior-net, the lightweight network---InDuDoNet w/o Prior-net (with 1,743,734 parameters) can still achieve the promising average PSNR/SSIM score which is comparable to the state-of-the-art baseline---DuDoNet++~\citep{lyu2020dudonet++} (with 25,983,627 parameters). This finely substantiates the claim that the full embedding of the inherent physical geometry constraints across the entire learning process can be helpful for the MAR task. {\textcolor{black}{From the STDs and boxplots, it can be seen that compared to InDuDoNet w/o Prior-net, our proposed InDuDoNet has lower PSNR/SSIM STDs and achieves higher median values and larger minimum values, which validates the role of Prior-net in steadily boosting the MAR performance. Note that the analysis about DuDoNet++ and the boxplots for the original input image are provided in Fig.~\ref{fig:allbox} below.}} Besides, as listed in Table~\ref{tab:claim}, under the setting with large metallic implants, the InDuDoNet w/o Prior-net is largely inferior to DuDoNet++, which is caused by the lack of normalization operation. These comprehensive results finely substantiate the
necessity and the rationality of carefully designing Prior-net in the next section.


\begin{figure*}[t]
	\begin{center}
		  \vspace{2mm}
		\includegraphics[width=0.7\linewidth]{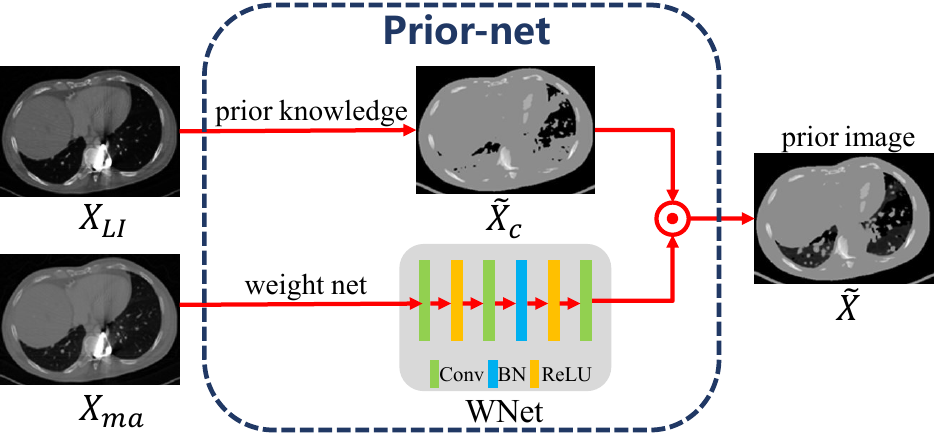}
	\end{center}
	\vspace{-4mm}
	\caption{\textcolor{black}{Knowledge-driven} Prior-net of InDuDoNet+, where $\odot$ is the element-wise multiplication. Our proposed Prior-net aims to learn the prior segmentation image $\widetilde{X}$ for obtaining the normalization coefficient $\widetilde{Y}$ via forward projection as $\widetilde{Y}=\mathcal{P}\widetilde{X}$.}
	\label{figwnet}
\end{figure*}

\subsection{\textcolor{black}{Knowledge-Driven Prior-net}}\label{sec:priordesign}
Although Prior-net is helpful for the MAR task, the previous InDuDoNet simply builds Prior-net upon the off-the-shelf U-shape network structure, which results in a relatively weak interpretability and does not fully consider the prior knowledge underlying the MAR task. Such a design leads to the difficulty for network module analysis and further performance improvement. Inspired by traditional MAR methods~\citep{meyer2010normalized}, we propose a novel knowledge-driven Prior-net. Concretely, the previous method~\citep{meyer2010normalized} adopted the thresholding-based hand-crafted design to segment the metal-corrupted CT image $X_{ma}$ and then captured the prior image $\widetilde{X}$ for CT reconstruction. The captured prior image $\widetilde{X}$ benefits the metal artifact reduction, since it takes the prior-knowledge, \emph{i.e.,} CT values among different tissues (\emph{e.g.}, low-density tissues and bones) are obviously different, into consideration. However, the generation process has a limitation---the thresholding-based segmentation strategy is sensitive to CT values.




{\textcolor{black}{To deal with the drawback, the proposed Prior-net first utilizes the thresholding-based clustering operator ~\citep{meyer2010normalized} to generate a coarse prior segmentation image $\widetilde{X}_{c}$, as shown in Fig.~\ref{figwnet}. Specifically, 
we first execute the k-means clustering on the artifact-reduced CT image $X_{LI}$ and then automatically get the segmentation thresholds. Following~\cite{meyer2010normalized}, we smooth $X_{LI}$ with a Gaussian filter for further artifact removal and then segment it into air, soft tissue, and bone via a simple thresholding. Based on the prior knowledge about the CT value distribution among different tissues, the air regions are then set to -1000 Hounsfield Units (HU), the soft tissue parts to 0~HU. Bone pixels keep their values, as they vary too much to properly model them with one value. The value that is assigned to metal is arbitrary since it does not affect the normalization and only the sinogram parts close to, but not inside, the metal trace contribute to the reconstructed result. Under this empirical knowledge based hand-crafted thresholding setting, air, soft tissue, and bone can be coarsely segmented, and then $\widetilde{X}_{c}$ is obtained. Considering that such hand-crafted thresholding based segmentation strategy may not be very accurate, we further refine $\widetilde{X}_{c}$ in a pixel-wise manner via a  weight matrix generated by a shallow WNet, only containing three convolutional layers, which results in a fine prior image $\widetilde{X}$ for the subsequent MAR task.}}

{\textcolor{black}{As seen, our proposed Prior-net is related to prior knowledge, which relies on the empirical knowledge to accomplish the extraction of $\widetilde{X}_{c}$. Clearly, this design is easy to understand: 1) the prior knowledge helps the coarse estimation; 2) the shallow CNN executes a more flexible adjustment. Besides, the role of network module involved in the new Prior-net can be easily evaluated (see Sec.~\ref{sec:abla}).}}

\vspace{3mm}
{\color{black}{\noindent\emph{\textbf{Remark:}} It should be noted that compared with other heuristic manners which adopt a general U-shape structure as a prior network, such as the PriorNet in DSCMAR~\citep{yu2020deep}, and the Prior-net in InDuDoNet~\citep{wang2021indudonet}, the proposed Prior-net in InDuDoNet+ has specific merits: 1)  The physical prior characteristics of tissues are finely integrated into the new Prior-net via $\widetilde{X}_{c}$, which is like a joint model-driven and data-driven manner and makes the entire network structure more transparent; 2) The WNet has a simple architecture, which significantly reduces the network parameters of InDuDoNet+ as well as improves its inference efficiency. The fewer network parameters have the potential to alleviate the overfitting problem and thereby improve the model generalization (see Sec.~\ref{sec:exp}); 3) Our Prior-net proposes a new method to embed the prior knowledge, \emph{i.e.}, CT values vary from different tissues, into deep networks, which will provide insights for the future research in this MAR field. It should be noted that as derived above, our proposed Prior-net aims to learn the normalized coefficient and it is an indispensable optimization part of the entire network structure which corresponds to an entire iterative algorithm. Hence, the proposed Prior-net cannot be regarded as an independent network module and directly applied to other methods, such as DSCMAR~\citep{yu2020deep}.}}

\section{Network Implementation}\label{sec:impdetail}
In this section, we present the implementation details of the proposed InDuDoNet+, including channel-wise concatenation and detachment operations, residual structures of {\normalsize{$\text{proxNet}_{\theta_{\widetilde{s}}^{(n)}}(\cdot)$}} and {\normalsize{$\text{proxNet}_{\theta_{x}^{(n)}}(\cdot)$}}, variable initialization ($\widetilde{S}_{0}$ and $X_{0}$), and training loss.

\begin{figure*}[t]
  \begin{center}
     \includegraphics[width=1\linewidth]{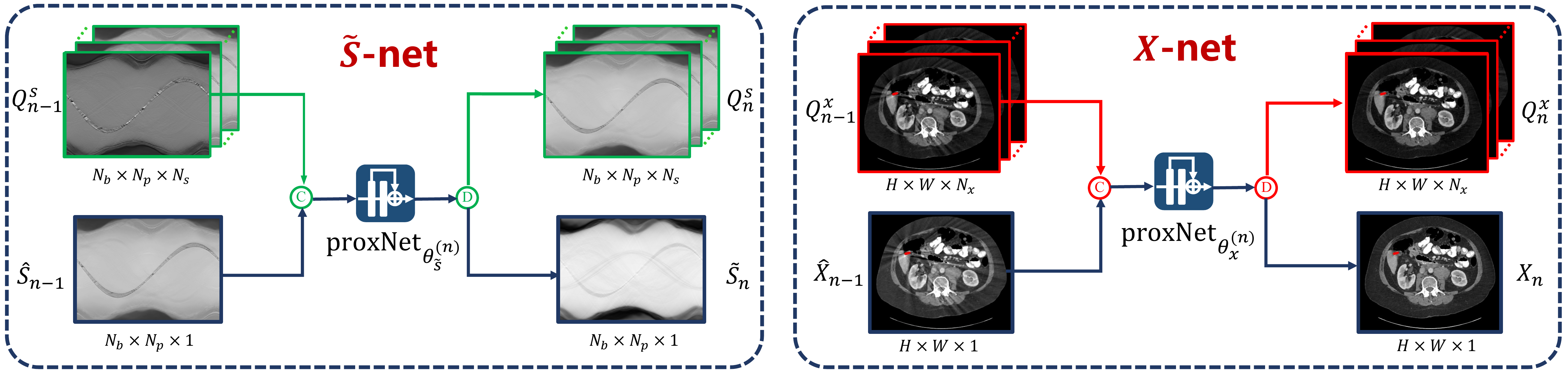}
  \end{center}
  \vspace{-1mm}
     \caption{Channel-wise concatenation and detachment operations involved in $\widetilde{S}$-net and $X$-net. Here $Q^s_{n}$ and $Q^x_{n}$ are auxiliary variables in sinogram domain and CT image domain, respectively, only for channel expansion.}
  \label{miafigdual}
\end{figure*}

\vspace{3mm}
\noindent\textbf{Channel-wise Concatenation and Detachment.} As shown in Fig.~\ref{fignet} (b), the images sent to {{$\text{proxNet}_{\theta_{\widetilde{s}}^{(n)}}(\cdot)$}} in $\widetilde{S}$-net and {{$\text{proxNet}_{\theta_{x}^{(n)}}(\cdot)$}} in $X$-net are $\widehat{S}_{n-1}$ and $\widehat{X}_{n-1}$, respectively, which are both gray images with a single channel. Such a single-channel input may be insufficient for deep networks to convey previous updating information for the iterations of $\widetilde{S}_{n}$ and $X_{n}$. For attaining possibly efficient information propagation, we impose the channel-wise concatenation and detachment operations~\citep{wang2020model} on the original $\widetilde{S}$-net and $X$-net shown in Fig.~\ref{fignet} (b).

Taking $\widetilde{S}$-net as an example, as shown in Fig.~\ref{miafigdual}, we additionally introduce a sinogram domain based auxiliary variable $Q^s_{n-1}\in \mathbb{R}^{N_{b}\times N_{p}\times N_{s}}$ for $\widetilde{S}$-net and concatenate it with the original input $\widehat{S}_{n-1}$ along the channel direction. The concatenated result is adopted as the new input for {{$\text{proxNet}_{\theta_{\widetilde{s}}^{(n)}}(\cdot)$}}, whose input dimension has been expanded from $N_{b}\times N_{p} \times 1$ to $N_{b}\times N_{p} \times (1+N_{s})$. Correspondingly, the output of {{$\text{proxNet}_{\theta_{\widetilde{s}}^{(n)}}(\cdot)$}} is with the size of $N_{b}\times N_{p} \times (1+N_{s})$. We divide it into two components along the channel direction, \emph{i.e.}, the first channel as the updated sinogram data $\widetilde{S}_{n}$ and the remaining channels as the updated auxiliary variable $Q^s_{n}$. Similar operations are executed on $X$-net.

\vspace{3mm}
\noindent\textbf{{{$\text{proxNet}_{\theta_{\widetilde{s}}^{(n)}}(\cdot)$}} and {{$\text{proxNet}_{\theta_{x}^{(n)}}(\cdot)$}}.} The {{$\text{proxNet}_{\theta_{\widetilde{s}}^{(n)}}(\cdot)$}} and {{$\text{proxNet}_{\theta_{x}^{(n)}}(\cdot)$}} in Fig.~\ref{miafigdual} have the same residual structure---four [{\normalsize{\emph{Conv+BN+ReLU+Conv+BN+Skip Connection}}}] residual blocks~\citep{he2016deep} at every stage, to represent the proximal operators $\mbox{prox}_{\lambda_{1}\eta_{1}}(\cdot)$ and  $\mbox{prox}_{\lambda_{2}\eta_{2}}(\cdot)$, respectively. The kernel size in every convolution layer is set to $3\times3$ with a stride of 1 and $N_{s}=N_{x}=32$. Actually, the effectiveness of adopting ResNet to describe a proximal operator has been fully verified by many existing studies for other computer vision tasks, such as spectral image fusion~\citep{xie2020mhf} and deraining~\citep{wang2020model}.


\vspace{3mm}
\noindent\textbf{Variable Initialization.}
To execute the iterative process, the variables $\widetilde{S}_{0}$, $X_{0}$, $Q^s_{0}$, and $Q^x_{0}$ first need to be initialized. By adopting the channel-wise concatenation and detachment operators, we initialize these variables as:
\begin{equation}\label{miccaiinis}
    \begin{split}
    &\{\widetilde{S}_{0} ~| ~Q^s_{0}\} = \mbox{proxNet}_{\theta_{\widetilde{S}}^{(0)}}\left(\text{concat}\left(Y_{LI},\mathcal{K}_{s}\otimes Y_{LI} \right)\right),\\
    &\{X_{0} ~| ~Q^x_{0}\} = \mbox{proxNet}_{\theta_{x}^{(0)}}\left(\text{concat}\left(X_{LI},\mathcal{K}_{x}\otimes X_{LI} \right)\right),
    \end{split}
\end{equation}
where `$|$' and `$\text{concat}(\cdot)$' represent the aforementioned channel-wise detachment and concatenation operation, respectively; $Y_{LI}$ and $X_{LI}$ are the reconstructed sinogram and CT images based on the traditional linear interpolation (LI) based method~\citep{kalender1987reduction}; $\mathcal{K}_{s}$ and $\mathcal{K}_{x}$ are the learnable convolutional filters with the size as ${f_{s} \times f_{s}\times N_{s}\times 1 }$ and ${f_{x} \times f_{x}\times N_{x}\times 1}$, respectively (in our experiments, ${f_{s} \times f_{s}\times N_{s}\times 1 }={f_{x} \times f_{x}\times N_{x}\times 1}=3\times 3 \times 32 \times 1$); $\otimes$ is the convolutional operator, which can be easily achieved by the current popular deep learning (DL) toolbox, such as Tensorflow\footnote{\url{https://tensorflow.google.cn/}} and PyTorch.\footnote{\url{https://pytorch.org/docs/stable/index.html}} $\mbox{proxNet}_{\theta_{\widetilde{S}}^{(0)}}$ and $\mbox{proxNet}_{\theta_{x}^{(0)}}(\cdot)$ are both ResNets with the same structures to {{$\text{proxNet}_{\theta_{\widetilde{s}}^{(n)}}(\cdot)$}} and  {{$\text{proxNet}_{\theta_{x}^{(n)}}(\cdot)$}}, respectively. Note that $\mbox{proxNet}_{\theta_{\widetilde{S}}^{(0)}}$ and $\mbox{proxNet}_{\theta_{x}^{(0)}}(\cdot)$ are trained together with {{$\text{proxNet}_{\theta_{\widetilde{s}}^{(n)}}(\cdot)$}} and {{$\text{proxNet}_{\theta_{x}^{(n)}}(\cdot)$}} in an end-to-end manner.

\vspace{3mm}
\noindent\textbf{Training Loss.} For network training, we adopt the mean squared error (MSE) for the extracted sinogram $\widetilde{Y}\odot\widetilde{S}_{n}$ and the estimated CT image $X_n$ at every stage as the training objective function:
\begin{equation}\label{Loss}
  \mathcal{L} = \sum_{n=0}^{N}\beta_{n}\left\|X_n-X_{gt} \right\|_F^2\odot(1-M)+\gamma\left(\sum_{n=1}^{N}\beta_{n}\left\| \widetilde{Y}\odot\widetilde{S}_n - Y_{gt}\right\|_F^2\right),
\end{equation}
\normalsize
where $X_{gt}$ and $Y_{gt}$ are the ground truth CT image and metal-free sinogram, respectively; $M$ is the binary metal mask. We simply set $\beta_{N}=1$ to make the outputs at the final stage play a dominant role, and $\beta_{n}=0.1$ ($n=0,\cdots, N-1$) to supervise each middle stage. $\gamma$ is a hyperparamter to balance the weight of different losses, which is empirically set to 0.1 in the experiments.


\section{Experiments}\label{sec:exp}
In this section, we first provide the detailed description of the experimental setting and then evaluate the performance of the proposed InDuDoNet+ on three medical image datasets by comparing with the existing representative MAR methods.

\subsection{Details Description}\label{thre}

\vspace{3mm}
\noindent\textbf{Synthesized DeepLesion.}\label{sec:expadd}
Following the simulation settings in~\citep{yu2020deep}, we randomly select a subset from the DeepLesion~\citep{yan2018deep} to synthesize metal-corrupted CT images {\textcolor{black}{with the fan-beam geometry}}.\footnote{\textcolor{black}{As stated in our open repository~\url{https://github.com/hongwang01/InDuDoNet}, the CT imaging code is provided by the author of DSCMAR~\citep{yu2020deep}.}} The metal masks are from~\citep{zhang2018convolutional}, which contain 100 metallic implants with different shapes and sizes. We choose 1,000 clean CT images and 90 metal masks to synthesize the training samples, and pair the additional 200 CT images from 12 patients with the remaining 10 metal masks to generate 2,000 images for testing. The sizes of the 10 metallic implants for test data are [2061, 890, 881, 451, 254, 124, 118, 112, 53, 35] in pixels. Consistent to~\citep{lin2019dudonet,lyu2020dudonet++}, we simply put the adjacent sizes into one group for average MAR performance evaluation. We adopt the procedures widely used by existing studies~\citep{zhang2018convolutional,liao2019adn,lin2019dudonet,yu2020deep,lyu2020dudonet++} to simulate $Y$ and $X_{ma}$. Various effects are considered during the simulation of metal artifacts, including polychromatic X-ray, partial volume effect, beam hardening, and Possion noise. All the CT images are resized to $416\times 416$ pixels and 640 projection views are uniformly spaced in 360 degrees. The size of the resulting sinogram $N_{b}\times N_{p}$ is set to $641\times 640$.


\vspace{3mm}
\noindent\textbf{Synthesized Dental.}
\textcolor{black}{To evaluate the generalization performance under the cross-body-site setting, we additionally collect several dental CT images~\citep{yu2020deep} and synthesize the corresponding metal-affected dental CT images according to the same simulation protocol executed on DeepLesion for performance evaluation.}

\vspace{3mm}
\noindent\textbf{Clinical SpineWeb.}
Furthermore, we evaluate the clinical feasibility of the proposed InDuDoNet+ using a clinical dataset, \emph{i.e.}, SpineWeb.\footnote{\url{http://spineweb.digitalimaginggroup.ca/Index.php?n=Main.Datasets}\label{spineweb}} Similar to \cite{liao2019adn}, we select the vertebrae localization and identification dataset from SpineWeb, which contains many CT images with metallic implants. Following the pre-processing protocol~\citep{liao2019adn}, we get metal-corrupted CT images for testing. The clinical images are resized and processed by using the same protocol to the synthesized data. Consistent to~\citep{liao2019adn,yu2020deep}, the clinical metal masks are segmented with a thresholding of 2500 Hounsfield Units (HU).


\vspace{3mm}
\noindent\textbf{Evaluation Metrics.}  For synthesized data, we adopt the peak signal-to-noise ratio (PSNR) and structured similarity index (SSIM) for quantitative evaluation. For clinical data, we only provide visual results due to the lack of ground truth CT images.

\vspace{3mm}
{\color{black}{\noindent\textbf{Training Details.} We implement our networks (\emph{i.e.},  InDuDoNet and InDuDoNet+) with PyTorch~\citep{paszke2017automatic} and differential operations $\mP$ and $\mP^T$ in the ODL library\footnote{\url{https://github.com/odlgroup/odl}.}} on an NVIDIA Tesla V100-SMX2 GPU.} The Adam optimizer with ($\beta_{1}$, $\beta_{2}$)=(0.5, 0.999) is adopted for network optimization. The initial learning rate is $2\times10^{-4}$ and divided by 2 every 40 epochs. The number of training epochs is 100 with a batch size of 1. Similar to~\citep{yu2020deep}, in each training iteration, we randomly select a clean CT image from the pool consisting of 1,000 images and a metal mask from the pool with 90 masks to synthesize a metal-affected sample. Following the previous InDuDoNet~\citep{wang2021indudonet}, we set the number of the total iterative stages $N$ to 10 \textcolor{black}{(see Sec.~\ref{sec:abla} for more analysis)}.

\vspace{3mm}
\noindent\textbf{Comparison Methods.}  We compare the proposed InDuDoNet+ with current state-of-the-art (SOTA) MAR approaches, including traditional LI~\citep{kalender1987reduction} and NMAR~\citep{meyer2010normalized}, deep learning (DL)-based CNNMAR~\citep{zhang2018convolutional}, DuDoNet~\citep{lin2019dudonet}, DSCMAR~\citep{yu2020deep}, DuDoNet++~\citep{lyu2020dudonet++} and our previous InDuDoNet \citep{wang2021indudonet}. For LI, NMAR, CNNMAR, and InDuDoNet, we directly use the released code. While for DuDoNet, DSCMAR, and DuDoNet++, we re-implement them since there is no official code.


\begin{table}[!t]
\centering
\footnotesize
\caption{PSNR (dB) /SSIM of different methods on synthesized DeepLesion. {\textcolor{black}{{\textcolor{black}{The column ``Average'' represents the PSNR/SSIM averagely computed on the entire dataset. The column ``STD'' denotes the standard derivation about the PSNR/SSIM results computed on the entire synthesized DeepLesion dataset.}} }} {\textcolor{black}{The column ``RMSE'' denotes the average root mean square error (RMSE) with unit HU.}} Bold and underline indicate the best and the second best results, respectively. \textcolor{black}{{$\text{DSCMAR}^{*}$}represents the results reported in the original work~\citep{yu2020deep}.}}\vspace{0mm}
\setlength{\tabcolsep}{2pt}
\begin{tabular}{l|c|c|c|c|c|c|c|c}
\Xhline{0.6pt}
Methods    & \multicolumn{5}{c|}{ Large Metal \quad \quad   \quad\quad  $\longrightarrow$    \quad   \quad\quad \quad         Small Metal}                & $\uparrow $Average  &\textcolor{black}{STD}  &{\textcolor{black}{$\downarrow $RMSE}}  \\
\Xhline{0.6pt}
{\textcolor{black}{Input}}             &{\textcolor{black}{24.12/0.6761}}              &{\textcolor{black}{26.13/0.7471}}              &{\textcolor{black}{27.75/0.7659}}               &{\textcolor{black}{28.53/0.7964}}              &{\textcolor{black}{28.78/0.8076}}              &{\textcolor{black}{27.06/0.7586}}  &\textcolor{black}{2.78/0.0610} &{\textcolor{black}{73.33HU}}\\
LI~\citep{kalender1987reduction}           &27.21/0.8920              &28.31/0.9185              &29.86/0.9464              &30.40/0.9555              &30.57/0.9608              &29.27/0.9347 &\textcolor{black}{3.90/0.0329} &{\textcolor{black}{53.37HU}}\\
NMAR~\citep{meyer2010normalized}              &27.66/0.9114              &28.81/0.9373              &29.69/0.9465              &30.44/0.9591              &30.79/0.9669              &29.48/0.9442   &\textcolor{black}{4.69/0.0634} &{\textcolor{black}{50.08HU}}          \\

CNNMAR~\citep{zhang2018convolutional}               &28.92/0.9433  &29.89/0.9588  & 30.84/0.9706             &31.11/0.9743              &31.14/0.9752              &30.38/0.9644   &\textcolor{black}{4.54/0.0165}           &{\textcolor{black}{45.69HU}} \\
DuDoNet~\citep{lin2019dudonet}            & 29.87/0.9723 & 30.60/0.9786 & 31.46/0.9839  & 31.85/0.9858 & 31.91/0.9862 & 31.14/0.9814 &\textcolor{black}{5.80/0.0116} &{\textcolor{black}{43.96HU}}\\
DSCMAR~\citep{yu2020deep}          & 34.04/0.9343 & 33.10/0.9362 & 33.37/0.9384  & 32.75/0.9393 & 32.77/0.9395 & 33.21/0.9375 &\textcolor{black}{4.04/0.0113} &{\textcolor{black}{32.16HU}}\\
\textcolor{black}{{$\text{DSCMAR}^{*}$}}         &{\textcolor{black}{- / -}} & {\textcolor{black}{- / -}} & {\textcolor{black}{- / -}} & {\textcolor{black}{- / -}} & {\textcolor{black}{- / -}} & \quad ~ {\textcolor{black}{- /0.9784}} & {\textcolor{black}{- / -}}  & {\textcolor{black}{{~31.15HU}}}\\
DuDoNet++~\citep{lyu2020dudonet++}         & 36.17/\underline{0.9784} & {38.34}/\underline{0.9891} & {40.32}/{0.9913}  & 41.56/0.9919 & {42.08}/{0.9921} & {39.69}/{0.9886} &\textcolor{black}{3.86/0.0075} &{\textcolor{black}{27.78HU}}\\
InDuDoNet~\citep{wang2021indudonet} &\textbf{36.74}/\textbf{0.9801} & \textbf{39.32}/\textbf{0.9896} & \textbf{41.86}/\underline{0.9931} & \underline{44.47}/\underline{0.9942} &\underline{45.01}/\underline{0.9948} &\underline{41.48}/\textbf{0.9904} &\textcolor{black}{4.73/0.0088}  &{\textcolor{black}{{\underline{18.20HU}}}}\\
InDuDoNet+ & \underline{36.28}/{0.9736} & \underline{39.23}/{0.9872} & \underline{41.81}/\textbf{0.9937} & \textbf{45.03}/\textbf{0.9952} & \textbf{45.15}/\textbf{0.9959}& \textbf{41.50}/\underline{0.9891} &\textcolor{black}{4.37/0.0066} &{\textcolor{black}{\textbf{16.93HU}}}\\ 
\Xhline{0.6pt}
\end{tabular}
\vspace{-1mm}
\label{tabsyn}
\end{table}
\normalsize

\begin{figure*}[t]
	{{\begin{center}
		{{\includegraphics[width=1.02\linewidth]{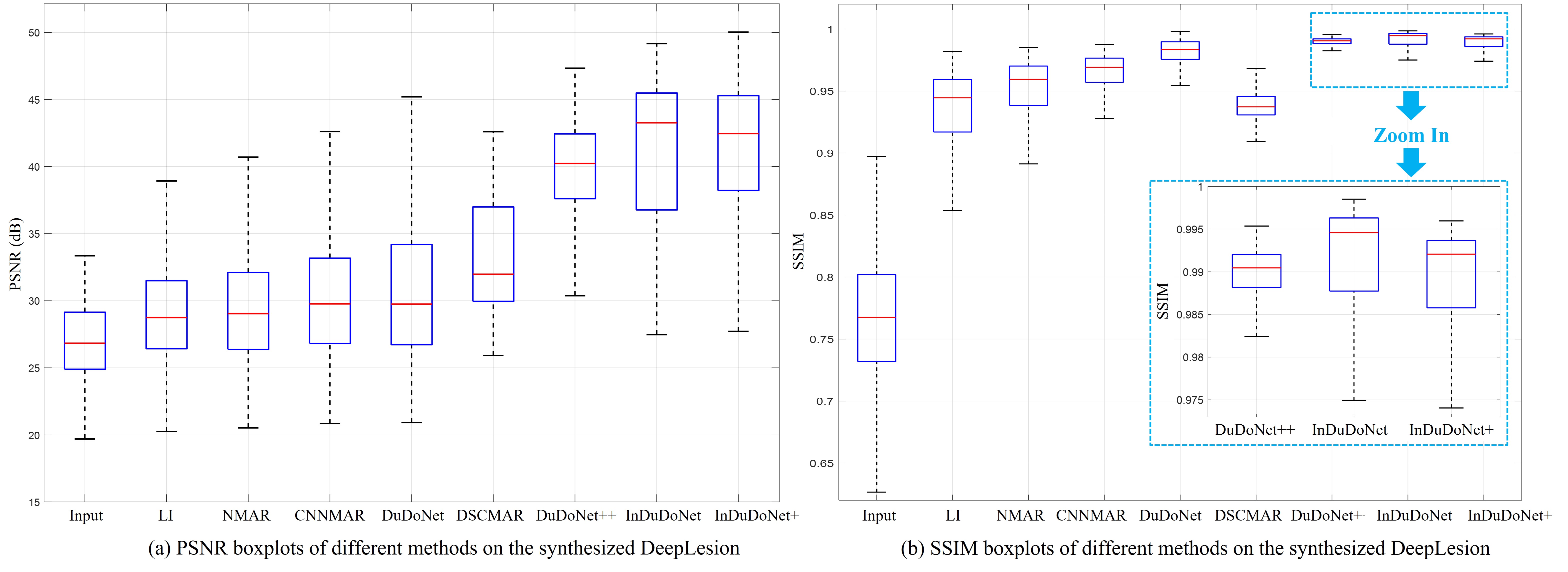}}}
	\end{center}
	\vspace{-6mm}
	\caption{{\textcolor{black}{Boxplots for the quantitative evaluation results of all the comparing methods in Table~\ref{tabsyn} on the synthesized DeepLesion dataset.}}}
	\label{fig:allbox}}}
\end{figure*}
\begin{figure*}[!t]
  \begin{center}
     \includegraphics[width=1\linewidth]{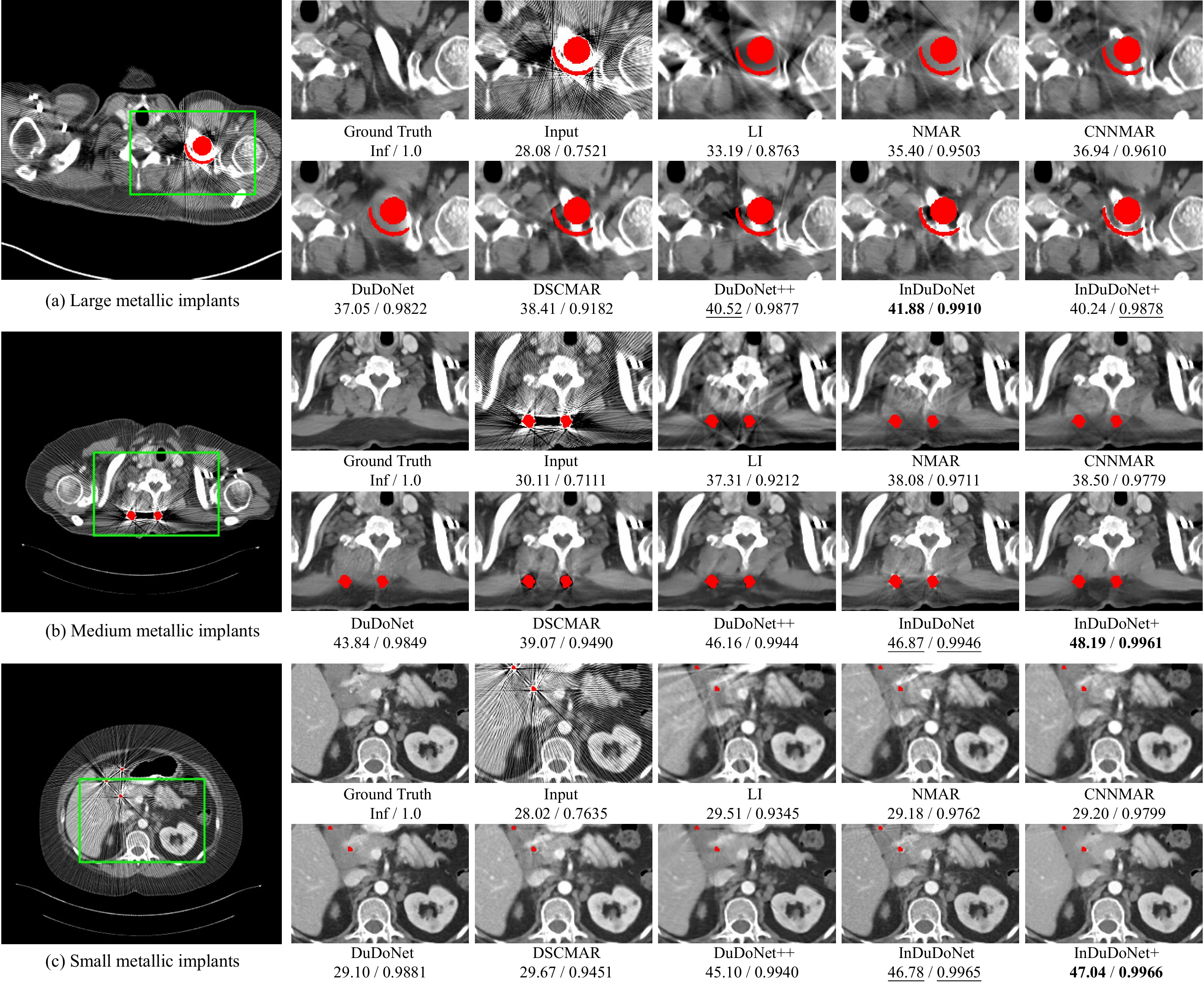}
  \end{center}
  \vspace{-6mm}
     \caption{Comparison of different MAR methods on the synthesized DeepLesion dataset with metallic implants of various sizes. PSNR (dB)/SSIM below is for reference. The display window is [-175, 275] HU. The red pixels stand for metallic implants.}
  \label{figsyn}
    \vspace{1mm}
\end{figure*}

\subsection{Experiments on Synthesized DeepLesion}

\vspace{3mm}
\noindent\textbf{Quantitative Comparison.}
Table~\ref{tabsyn} reports the quantitative comparison of different MAR methods on synthesized DeepLesion. We can observe that most of DL-based methods consistently outperform the conventional LI and NMAR \textcolor{black}{with higher PSNR/SSIM and lower RMSE}, showing the superiority of data-driven deep CNN for MAR. The dual enhancement approaches (\emph{i.e.,} DuDoNet, DSCMAR, and DuDoNet++) achieve higher PSNR than the sinogram-enhancement-only CNNMAR. Compared with DuDoNet, DSCMAR, and DuDoNet++, our dual-domain methods (\emph{i.e.}, InDuDoNet and InDuDoNet+) explicitly embed the physical CT imaging geometry constraints into the mutual learning between spatial and Radon domains, \emph{i.e.,} jointly regularizing the sinogram and CT image recovered at each stage. Hence, our methods achieve the most competitive PSNRs for all metal sizes as listed. {\textcolor{black}{Fig.~\ref{fig:allbox} shows the boxplots for the PSNR/SSIM results of different comparing methods on the entire synthesized DeepLesion dataset. 
As seen, for our proposed methods, \emph{i.e.}, InDuDoNet and InDuDoNet+, the maximum values and the median values of PSNR/SSIM obviously outperform other comparing methods. Besides, we can find that the minimum values (corresponding to the large metal setting) of PSNR/SSIM of our proposed methods are slightly lower than that of DuDoNet++. This can be explained by the fact that the larger metallic implants, the more severely damaged images would be, making it more difficult to restore the image details. Hence, the room for consistent performance improvement under this large metal setting is limited.
The similar trend can also be observed from Table~\ref{tabsyn}. As seen, the proposed methods outperform DuDoNet++ with a large margin under the small metal setting, while slightly outperforming under large metal implants. Even so, our proposed methods can still obtain comparable PSNR/SSIM STDs and lower RMSE, showing good MAR performance.}}

{\textcolor{black}{To comprehensively substantiate the advantages of our proposed method, we also provide the p-value analysis. Specifically, for the competitive dual-domain based deep MAR methods, such as, DSCMAR, DuDoNet, DuDoNet+, InDuDoNet, and the proposed InDuDoNet+, the corresponding STD of the average PSNR is 0.0189, 0.0346, 0.0232, 0.0178, and 0.0156, respectively, which shows the better robustness of our InDuDoNet+.}} For the paired t-test comparing against InDuDoNet+, the p-value is 0.0342 for InDuDoNet, 0.0246 for DuDoNet++, and less than 0.001 for other baselines. As seen, all p-values are less than the significance level 0.05, which validates that the performance improvement achieved by our method over existing methods is significant.


\vspace{3mm}
\noindent\textbf{Visual Comparison.} The visual comparisons are shown in Fig.~\ref{figsyn}.
We find that although LI, NMAR, and CNNMAR can remove obvious streaky artifacts, they introduce secondary artifacts and lose useful image details to a certain extent, which is caused by the discontinuity in the corrected sinogram. DuDoNet and DuDoNet++ both produce over-smoothed artifact-removed image, which is mainly due to the lack of physical geometry constraint on the final output of image enhancement module.
{\color{black}{For DSCMAR, the image details can be preserved well and the
reconstructed CT images have better sharpness. This is mainly attributed to that the final recovered
CT image is obtained via the filtered back projection operation which can alleviate the over-smoothness problem caused by CNN. However, the image intensity is not very accurate, for example, for the bone structures. This is possibly because that the learnable flexibility of DSCMAR is reduced since the final CT image is directly reconstructed in a fixed
analytic manner, and the prior image is not sufficiently accurate for sinogram completion, which would largely affect the contrast of the artifact-reduced CT image.}} Comparatively, our methods not only evidently remove more artifacts but also better preserve the image details.

\begin{figure*}[!t]
  \begin{center}
  \vspace{-3mm}
     \includegraphics[width=1\linewidth]{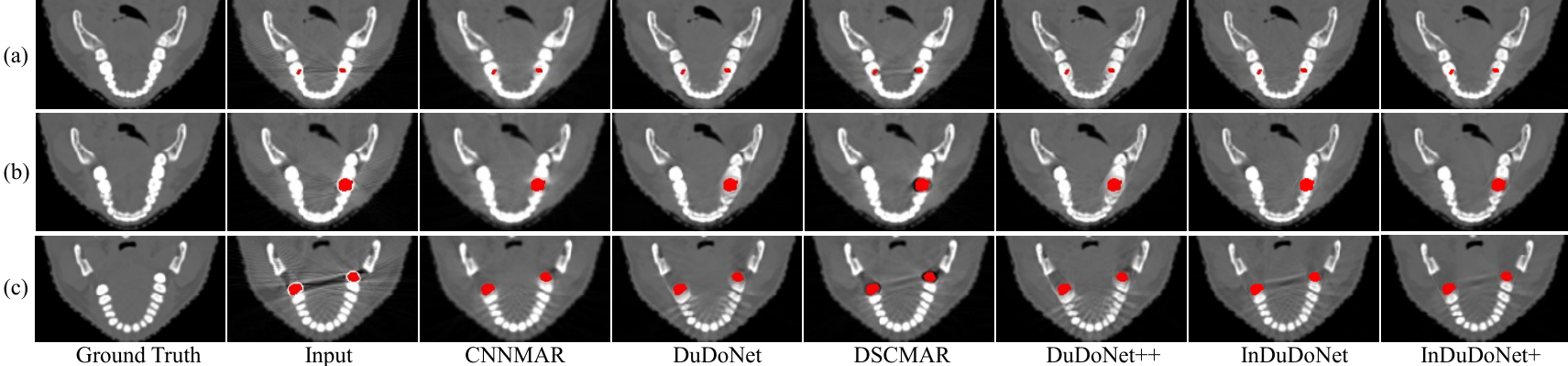}
  \end{center}
  \vspace{-6mm}
     \caption{Performance comparison on the synthesized Dental dataset with different numbers of dental fillings where all the DL-based MAR methods are trained on synthesized DeepLesion.}
  \label{figdental}
    \vspace{-2mm}
\end{figure*}

\begin{table}[!t]
\centering
\footnotesize
\caption{The first three rows represent the PSNR (dB)/SSIM of different MAR methods on synthesized Dental CT images shown in Fig.~\ref{figdental}. {\textcolor{black}{The row ``Whole'' represents the average PSNR/SSIM of different methods on the whole synthesized Dental CT dataset~\citep{yu2020deep}. The row ``STD'' represents the standard derivation about the PSNR/SSIM results computed on the entire synthesized Dental dataset.}}}
\setlength{\tabcolsep}{3pt}
\begin{tabular}{c|c|c|c|c|c|c|c|c|c}
\Xhline{0.6pt}
Figure        & Input & LI            &NMAR    &CNNMAR &DuDoNet & DSCMAR  & DuDoNet++ & InDuDoNet & InDuDoNet+     \\
\Xhline{0.6pt}
(a) & 37.08/0.9157& 33.98/0.9357  & 35.03/0.9569  & 36.65/0.9747   & 39.07/0.9753 & 37.14/0.9751 & 40.04/\underline{0.9900} & \underline{43.33}/0.9731 & \textbf{43.64}/\textbf{0.9922}\\ 
(b) & 36.46/0.9332 & 32.83/0.9217  & 33.57/0.9384 & 36.33/0.9690  & 38.09/0.9741  & 37.17/0.9784  & 39.16/\underline{0.9881}  &  \underline{42.61}/0.9727  & \textbf{43.01}/\textbf{0.9924} \\ 
(c) & 34.19/0.8733 & 33.62/0.9129  & 34.98/0.9523 & 36.61/0.9746 & 37.75/0.9747  & 37.15/0.9796 & 38.45/\underline{0.9883}  &  \underline{41.66}/0.9700 & \textbf{42.69}/\textbf{0.9894}\\
\hline
\hline
\textcolor{black}{Whole} & \textcolor{black}{35.34/0.9024} & \textcolor{black}{33.01/0.9140}  & \textcolor{black}{33.89/0.9386} & \textcolor{black}{36.27/0.9677} & \textcolor{black}{37.70/0.9732}  & \textcolor{black}{36.99/0.9770} & \textcolor{black}{38.51/\underline{0.9882}}  &  \textcolor{black}{\underline{41.95}/0.9757} & \textcolor{black}{\textbf{42.68}/\textbf{0.9910}} \\
\textcolor{black}{STD} & \textcolor{black}{1.51/0.0263} &\textcolor{black}{1.09/0.0242}  & \textcolor{black}{1.32/0.0231} &\textcolor{black}{0.82/0.0086} &\textcolor{black}{1.49/0.0025}  & \textcolor{black}{0.55/0.0035} & \textcolor{black}{1.85/0.0018}  &  \textcolor{black}{1.00/0.0044} & \textcolor{black}{1.16/0.0017} \\
\Xhline{0.6pt}
\end{tabular}
\label{tabdental}
\end{table}

\begin{figure*}[!t]
	{{\begin{center}
		{{\includegraphics[width=1.02\linewidth]{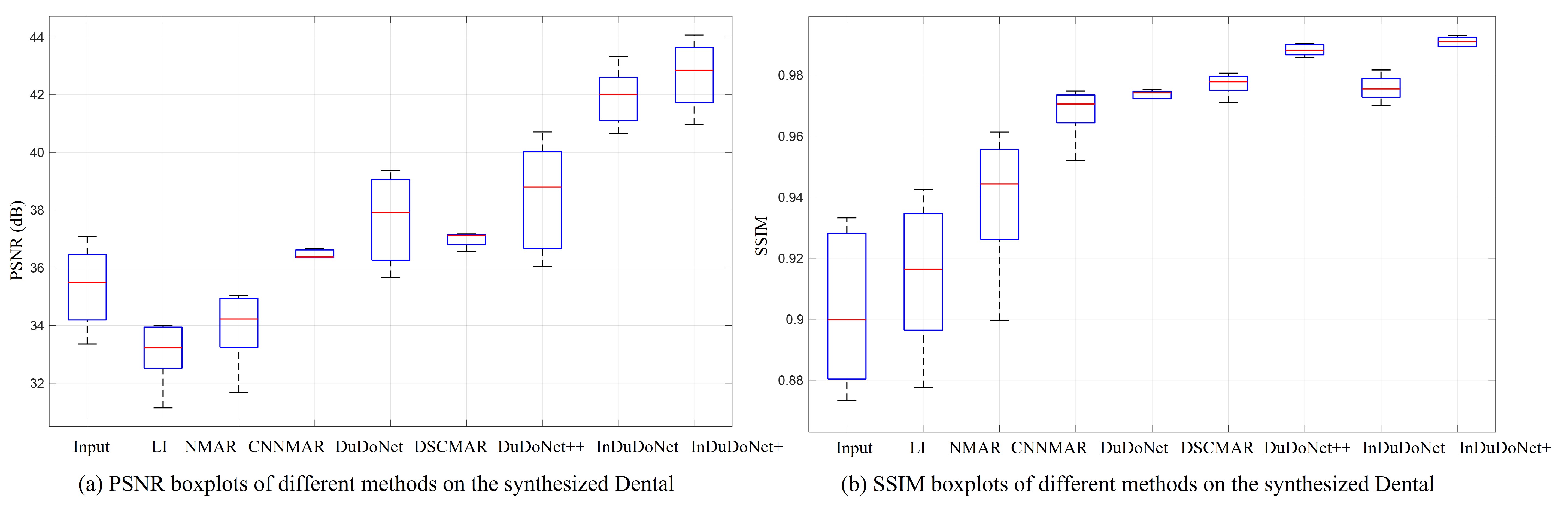}}}
	\end{center}
	\vspace{-6mm}
	\caption{{\textcolor{black}{Boxplots for the quantitative results of all the comparing methods in Table~\ref{tabdental} on the synthesized Dental CT dataset.}}}
	\label{fig:dentalbox}}}
\end{figure*}
\begin{figure*}[!t]
  \begin{center}
     \includegraphics[width=1\linewidth]{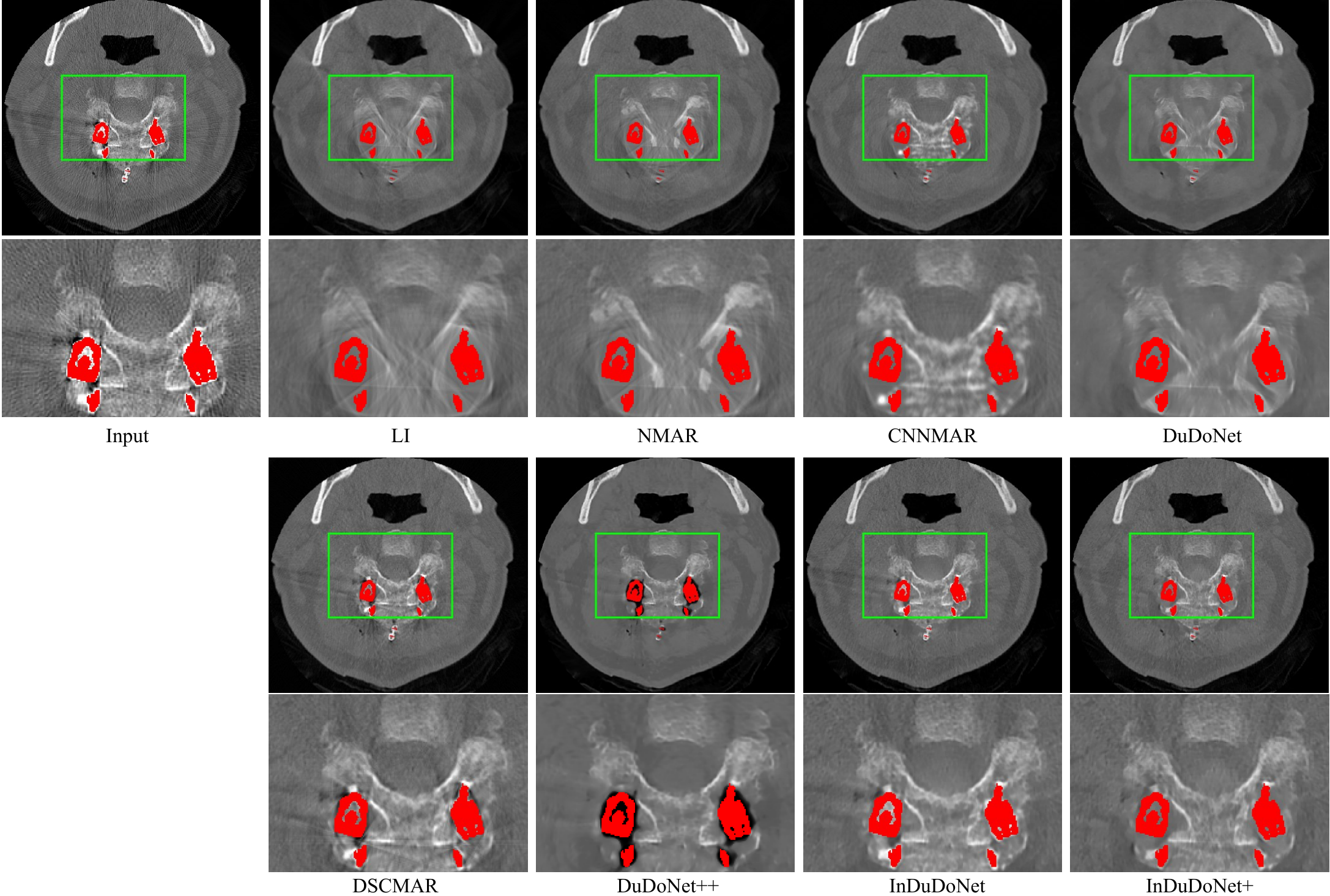}
  \end{center}
  \vspace{-5mm}
     \caption{Performance comparison on the clinical SpineWeb dataset. All the DL-based MAR methods are trained on synthesized DeepLesion. For each method, the first row is the generalized result and the second row is the zoomed view of the region marked with the green box for better visualization. The red pixels stand for metallic implants.}
  \label{figspine60}
    \vspace{-2mm}
\end{figure*}

\subsection{Generalization to Synthesized Dental}\label{sec:genera1}
Fig.~\ref{figdental} displays the visual comparison of different MAR methods on synthesized dental CT images with different numbers of dental fillings, where all the DL-based MAR comparison methods are trained on synthesized DeepLesion data (focusing on abdomen and thorax). {\textcolor{black}{The corresponding quantitative results are reported in Table~\ref{tabdental} where the average PSNR/SSIM and the STDs on the entire Dental datset are also included.}}

From the listed results, we have several observations: 1) Due to the domain gap between thorax CT and dental CT, almost all the benchmarking methods leave obvious artifacts in the reconstructed CT images to some extent. In contrast, the performances of our InDuDoNet+ and the previously proposed InDuDoNet are still competitive, owning to the inherent incorporation of physical imaging constraints; 2) In the cross-body-site (from abdomen/thorax CT to dental CT) scenario, InDuDoNet+ is evidently superior to InDuDoNet, which substantiates the effectiveness of the proposed model-driven Prior-net. {\textcolor{black}{Fig.~\ref{fig:dentalbox} presents the boxplots for PSNR/SSIM of different comparing methods on the entire synthesized Dental datasets. It can be seen that our proposed InDuDoNet+ steadily achieves higher PSNR/SSIM scores and shows competitive generalization ability.}}

\subsection{Generalization to Clinical SpineWeb}\label{sec:genera2}
We further evaluate all MAR methods on the clinical SpineWeb dataset. The experimental results are shown in Fig.~\ref{figspine60}. Due to the inaccurate sinogram completion, LI and NMAR introduce obvious secondary artifacts. CNNMAR, DuDoNet, and DuDoNet++ evidently blur the image details. DSCMAR fails to remove obvious dark shadings and streaky artifacts. For the existing MAR approaches, the degradation of MAR performance is mainly caused by the large domain gap between the synthesized DeepLesion (abdomen and thorax CT) and SpineWeb (spine CT). Compared with the previous InDuDoNet, our InDuDoNet+ removes more artifacts and preserves the image details better. This comparison finely verifies that the Prior-net tends to better regularize the network learning and thus improve its generalization performance. {\textcolor{black}{Note that since there is no ground truth for the publicly available clinical SpineWeb dataset, following~\cite{liao2019adn}, we can only provide the visual comparison.}}


\begin{figure*}[!t]
  \begin{center}
     \includegraphics[width=1\linewidth]{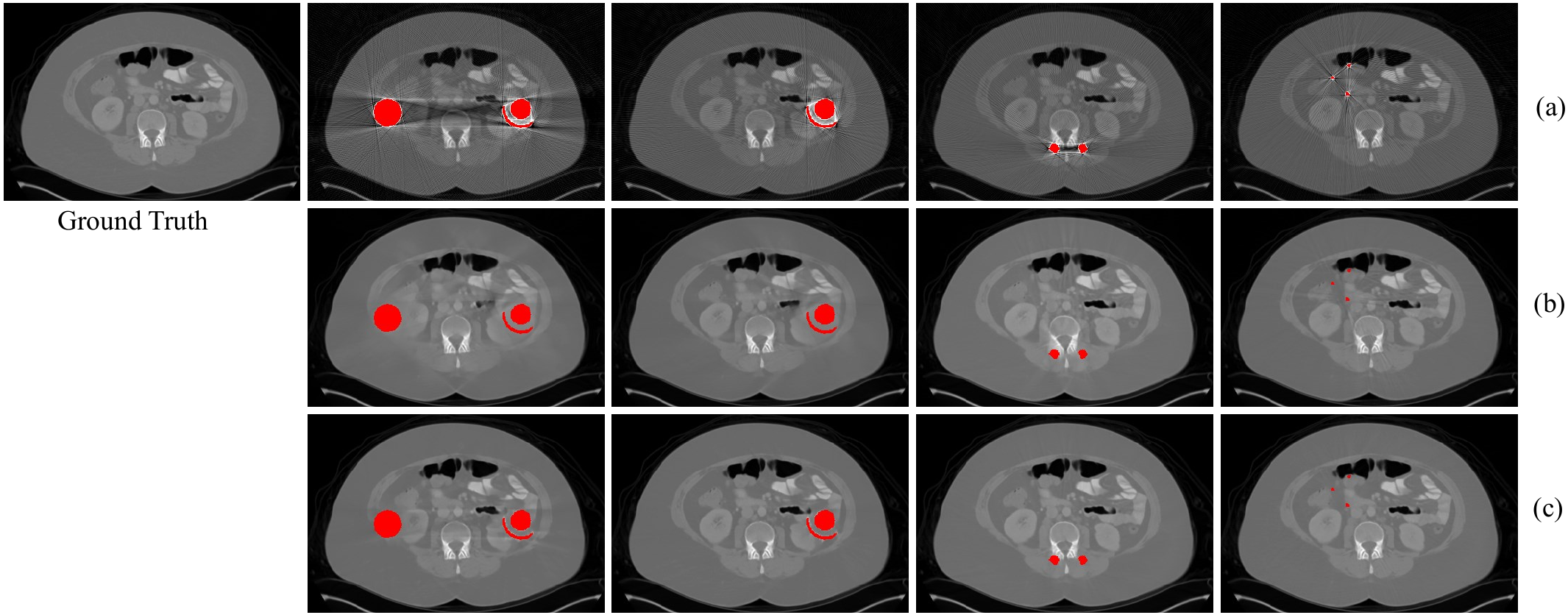}
  \end{center}
  \vspace{-4mm}
     \caption{ (a) Ground truth and metal-corrupted CT images with different metallic implants selected from synthesized DeepLesion; (b) The artifact-reduced results recovered by ``InDuDoNet+ w/o WNet''; (c) The images predicted by InDuDoNet+. The red pixels stand for metals.}
  \label{figablaplus}
    \vspace{1mm}
\end{figure*}

\begin{table}[!t]
\centering
\small
\caption{Effect of WNet on the performance of InDuDoNet+ on the synthesized DeepLesion dataset. {\textcolor{black}{The column ``Average'' represents the PSNR/SSIM averagely computed on the entire dataset. The column ``STD'' denotes the standard derivation about the PSNR/SSIM results computed on the entire dataset.}}}
\setlength{\tabcolsep}{3.1pt}
\begin{tabular}{l|c|c|c|c|c|c|c}
\Xhline{0.6pt}
Methods    & \multicolumn{5}{c|}{ Large Metal \quad \quad   \quad\quad  $\longrightarrow$    \quad   \quad\quad \quad         Small Metal}                & Average  & \textcolor{black}{STD}     \\
\Xhline{0.6pt}
{\textcolor{black}{Input}}             &{\textcolor{black}{24.12/0.6761}}              &{\textcolor{black}{26.13/0.7471}}              &{\textcolor{black}{27.75/0.7659}}               &{\textcolor{black}{28.53/0.7964}}              &{\textcolor{black}{28.78/0.8076}}              &{\textcolor{black}{27.06/0.7586}} &\textcolor{black}{2.78/0.0610}\\
InDuDoNet+ w/o WNet           & 31.07/0.9511 & 33.66/0.9682 & 37.65/0.9823  & 40.01/0.9864 & 40.36/0.9881 & 36.55/0.9752 &\textcolor{black}{5.02/0.0175}\\
InDuDoNet+ & {36.28}/{0.9736} & {39.23}/{0.9872} &{41.81}/{0.9937} &{45.03}/{0.9952} &{45.15}/{0.9959}& {41.50}/{0.9891} &\textcolor{black}{4.37/0.0066}\\ 
\Xhline{0.6pt}
\end{tabular}
\vspace{-1mm}
\label{tabablanet}
\end{table}
\normalsize
\begin{figure*}[!t]
	\begin{center}
		  \vspace{2mm}
		{{\includegraphics[width=0.75\linewidth]{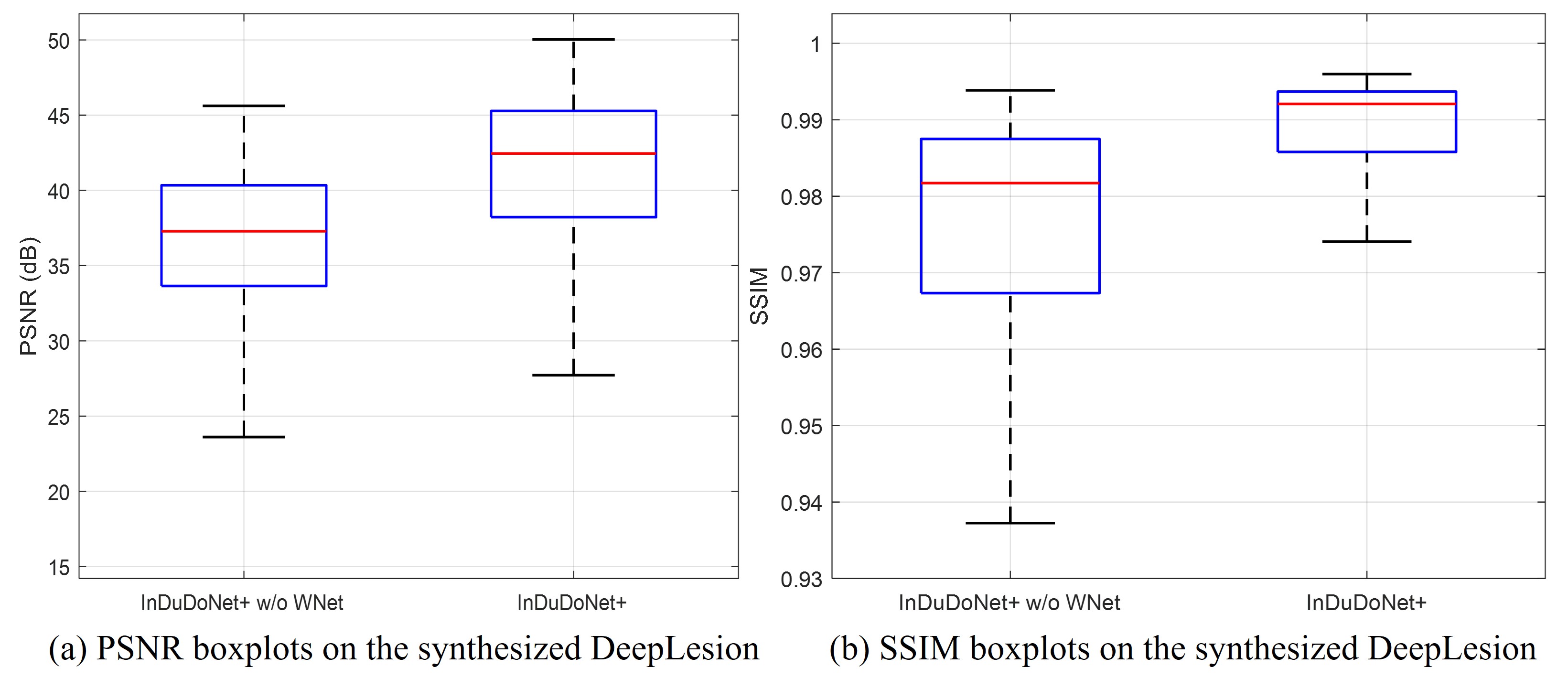}}}
	\end{center}
	\vspace{-6mm}
	\caption{{\textcolor{black}{Boxplots for the quantitative results of different methods in Table~\ref{tabablanet} on the synthesized DeepLesion dataset.}}}
	\label{fig:nowbox}
\end{figure*}


\subsection{Ablation Study}\label{sec:abla}
{\textcolor{black}{Here we further conduct a series of ablation studies to comprehensively analyze the effect of different factors on our InDuDoNet+, including network module, loss function, the number of iterative stages $N$, and clinical segmentation mask.}}


\vspace{3mm}
\noindent{\textcolor{black}{\textbf{Network Module}}.} Fig.~\ref{figablaplus} shows the visual comparison between ``InDuDoNet+ w/o WNet'' and InDuDoNet+ on the synthesized DeepLesion, where ``InDuDoNet+ w/o WNet'' means that we directly utilize the hand-crafted coarse prior segmentation image $\widetilde{X}_{c}$ as the final prior image $\widetilde{X}$, while omitting WNet in Fig.~\ref{figwnet}. {\textcolor{black}{Table~\ref{tabablanet} reports the quantitative evaluation and Fig.~\ref{fig:nowbox} illustrates the corresponding boxplots for the PSNR and SSIM distributions on the entire dataset.}} From the results, {\textcolor{black}{we can find that 1) In ``InDuDoNet+ w/o WNet'', the hand-crafted  segmentation manner would cause the large error in $\widetilde{Y}$, and then seriously affect the reconstruction performance of $S$ and largely impair the recovery quality of CT image $X$, leading to the severe performance degradation. Clearly, if the Prior-net is not properly designed, we cannot fully leverage the potential of the physical imaging mechanism in helping boost the MAR performance; 2) Due to the high learning flexibility of prior image, the embedding of WNet helps InDuDoNet+ achieve a significant MAR performance improvement. This finely validates the analysis in Sec.~\ref{sec:priordesign} that the introduction of WNet is helpful for refining the thresholding-based prior segmentation result and such data-driven adjustment manner is more flexible.}} {\textcolor{black}{It should be noted that we cannot perform an ablation study about the $\widetilde{S}$-net and $X$-net separately. As presented in Sec.~\ref{sec:details}, the proposed network structure (including prior-net, $\widetilde{S}$-net, and $X$-net) is correspondingly constructed based on an iterative optimization algorithm for solving the dual-domain model as Eq.~\eqref{o3}. Hence, $\widetilde{S}$-net and $X$-net are both an indispensable part of the entire InDuDoNet+.}}

\vspace{3mm}
\noindent{\textcolor{black}{\textbf{Loss Function.} Table~\ref{tab:loss} lists the average PSNR/SSIM on the synthesized DeepLesion where our proposed InDuDoNet+ is trained with different loss function settings. Specifically, from Variant 1, we can find that even with a single loss term on the final reconstructed CT image $X_{N}$, our method can still achieve the competitive performance and obtain a higher PSNR score than the state-of-the-art comparison method---DuDoNet+ (40.34~dB vs. 39.69~dB). By comparing Variants 1 and 3, we can see that the MAR performance of our proposed InDuDoNet+ can be further improved by imposing supervision loss on the final extracted sinogram $\widetilde{Y}\odot\widetilde{S}_{N}$. Besides, by correspondingly comparing Variants 1 and 2, it is obvious that adopting intra-stage supervision on the results $X_{n}$ is helpful for enabling the network to be evolved to a better direction and then achieve higher PSNR/SSIM, which demonstrates the rationality and the effectiveness of the proposed mutual iterative learning mechanism between sinogram data and CT image data. Based on this analysis, we thus present Variant 4 and take it as our final loss function in all comparison experiments where we directly set $\beta_{N}=1$ and $\gamma=0.1$ to make the outputs at the final stage play a dominant role, and $\beta_{n}$ ($n=0,\cdots, N-1$) as 0.1 to help find the correct parameter at each stage. From these results, we can also observe that under different hyper-parameter settings, our method can always achieve the satisfactory performance. Such fine results are mainly attributed to the theoretically well-founded design of the adopted deep unfolding network which can inherently push the network to optimize in a right direction.}}

\begin{table*}[t]
\centering
\caption{ {\color{black}{Average PSNR (dB)/SSIM on the synthesized DeepLesion dataset of InDuDoNet+ with different loss functions.}}}
\small
\setlength{\tabcolsep}{6pt}
 {\color{black}{\begin{tabular}{c|c|c|c}
\hline
Variant & Parameter setting & Loss function & Average\\

\hline
1 & \tabincell{c}{$\gamma=0$, $\beta_{n}=0~(n\neq N)$, $\beta_{N}=1$}  & $\mathcal{L}=\left\|X_N-X_{gt} \right\|_F^2\odot(1-M)$ &40.34/0.9884  \\
\hline
2 & \tabincell{c}{$\gamma=0$, $\beta_{n}=0.1~(n\neq N)$, $\beta_{N}=1$} & $\mathcal{L}= \left\|X_N-X_{gt} \right\|_F^2\odot(1-M) + \sum_{n=0}^{N-1}0.1\left\|X_n-X_{gt} \right\|_F^2\odot(1-M)$ &41.23/0.9889 \\
\hline
3 & \tabincell{c}{$\gamma=0.1$, $\beta_{n}=0~(n\neq N)$, $\beta_{N}=1$} & $\mathcal{L}= \left\|X_N-X_{gt} \right\|_F^2\odot(1-M) +0.1\left\| \widetilde{Y}\odot\widetilde{S}_N - Y_{gt}\right\|_F^2$ &41.31/0.9890\\
\hline
4 & \tabincell{c}{$\gamma=0.1$, $\beta_{n}=0.1~(n\neq N)$, $\beta_{N}=1$} & \tabincell{c}{$\mathcal{L}= \left\|X_N-X_{gt} \right\|_F^2\odot(1-M) +\sum_{n=0}^{N-1}0.1\left\|X_n-X_{gt} \right\|_F^2\odot(1-M)$\\ $+ 0.1\left(\left\| \widetilde{Y}\odot\widetilde{S}_N - Y_{gt}\right\|_F^2 +\sum_{n=1}^{N-1}0.1\left\| \widetilde{Y}\odot\widetilde{S}_n - Y_{gt}\right\|_F^2\right)$} &\textbf{41.50}/\textbf{0.9891}\\
\hline
\end{tabular}}}
\vspace{-3mm}
\label{tab:loss}
\end{table*}

\begin{table}[t]
\centering
\caption{ {\color{black}{Effect of the stage number $N$ on the performance of InDuDoNet+ on the synthesized DeepLesion, measured in PSNR and SSIM.}}}
\setlength{\tabcolsep}{9pt}
 {\color{black}{\begin{tabular}{l|c|c|c|c|c|c}
\Xhline{0.6pt}
   $N$  & \multicolumn{5}{c|}{ Large Metal \quad \quad   \quad\quad  $\longrightarrow$    \quad   \quad\quad \quad         Small Metal}                & Average\\
\Xhline{0.6pt}
$N$=0              &28.91/0.9280              &30.42/0.9400              &34.45/0.9599               &36.72/0.9653              &37.18/0.9673              &33.54/0.9521              \\
$N$=1              &30.26/0.9465              &32.86/0.9651              &36.97/0.9821              &39.50/0.9869              &39.95/0.9885              &35.91/0.9738              \\
$N$=3            &32.12/0.9657  &35.16/0.9784  &38.38/0.9871              &41.06/0.9900              &41.33/0.9911             &37.61/0.9824              \\
$N$=6          &34.32/\textbf{0.9758} & 37.14/0.9851 &40.99/0.9917  &43.33/0.9935 &43.55/0.9942 & 39.87/0.9881\\
$N$=8          &34.55/0.9742 & 37.01/0.9850 &\underline{41.85}/0.9923  &44.36/0.9943 &43.79/0.9944 &40.31/0.9881 \\
$N$=10      &\underline{36.28}/{0.9736} & \underline{39.23}/\underline{0.9872} &41.81/\underline{0.9937} & \textbf{45.03}/\underline{0.9952} & \underline{45.15}/\underline{0.9959}& \underline{41.50}/\underline{0.9891} 
\\
$N$=12   &\textbf{36.33}/\underline{0.9746} &\textbf{39.39}/\textbf{0.9884} &\textbf{42.05}/\textbf{0.9945} &\underline{45.02}/\textbf{0.9957} &\textbf{45.36}/\textbf{0.9962} &\textbf{41.63}/\textbf{0.9899}\\
\Xhline{0.6pt}
\end{tabular}}}
\label{tab:n}
\end{table}

\begin{figure*}[t]
	{\textcolor{black}{\begin{center}
		  \vspace{2mm}
		{\textcolor{black}{\includegraphics[width=1\linewidth]{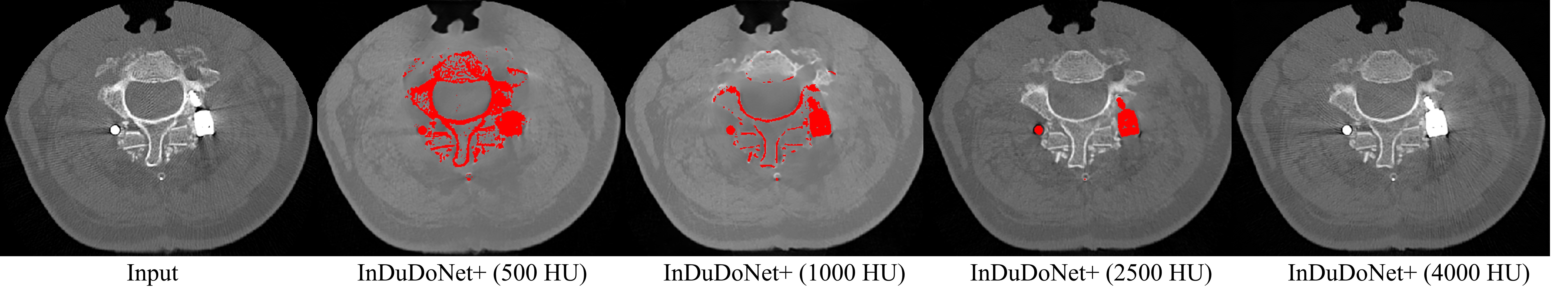}}}
	\end{center}
	\vspace{-6mm}
	\caption{{\textcolor{black}{Generalization performance of our proposed InDuDoNet+ on the clinical SpineWeb dataset where the clinical metal mask is segmented based on different thresholds, including 500~HU, 1000~HU, 2500~HU, and 4000~HU. The red pixels stand for metallic implants.}}}
	\label{fig:mask}}}
\end{figure*}

\begin{table}[t]
\centering
\small
\caption{{\color{black}{PSNR (dB) /SSIM of different ways to generate the artifact-reduced CT image from the proposed InDuDoNet+ on synthesized DeepLesion.}}}
\setlength{\tabcolsep}{8.2pt}
 {\color{black}{\begin{tabular}{l|c|c|c|c|c|c}
\Xhline{0.6pt}
Methods    & \multicolumn{5}{c|}{ Large Metal \quad \quad   \quad\quad  $\longrightarrow$    \quad   \quad\quad \quad         Small Metal}                & Average      \\
\Xhline{0.6pt}
Input             &24.12/0.6761              &26.13/0.7471              &27.75/0.7659               &28.53/0.7964              &28.78/0.8076              &27.06/0.7586              \\
$\text{InDuDoNet+}_{{fbp}}$ &32.78/0.9043 &35.19/0.9188 &37.93/0.9278 &38.90/0.9306 &38.95/0.9309 &36.75/0.9225\\
$\text{InDuDoNet+}$ &36.28/0.9736 &39.23/0.9872 &41.81/0.9937 &45.03/0.9952&45.15/0.9959&41.50/0.9891\\
\Xhline{0.6pt}
\end{tabular}}}
\label{tab:sx}
\end{table}
\normalsize

\begin{figure*}[t]
	\begin{center}
		  \vspace{2mm}
		\includegraphics[width=1.04\linewidth]{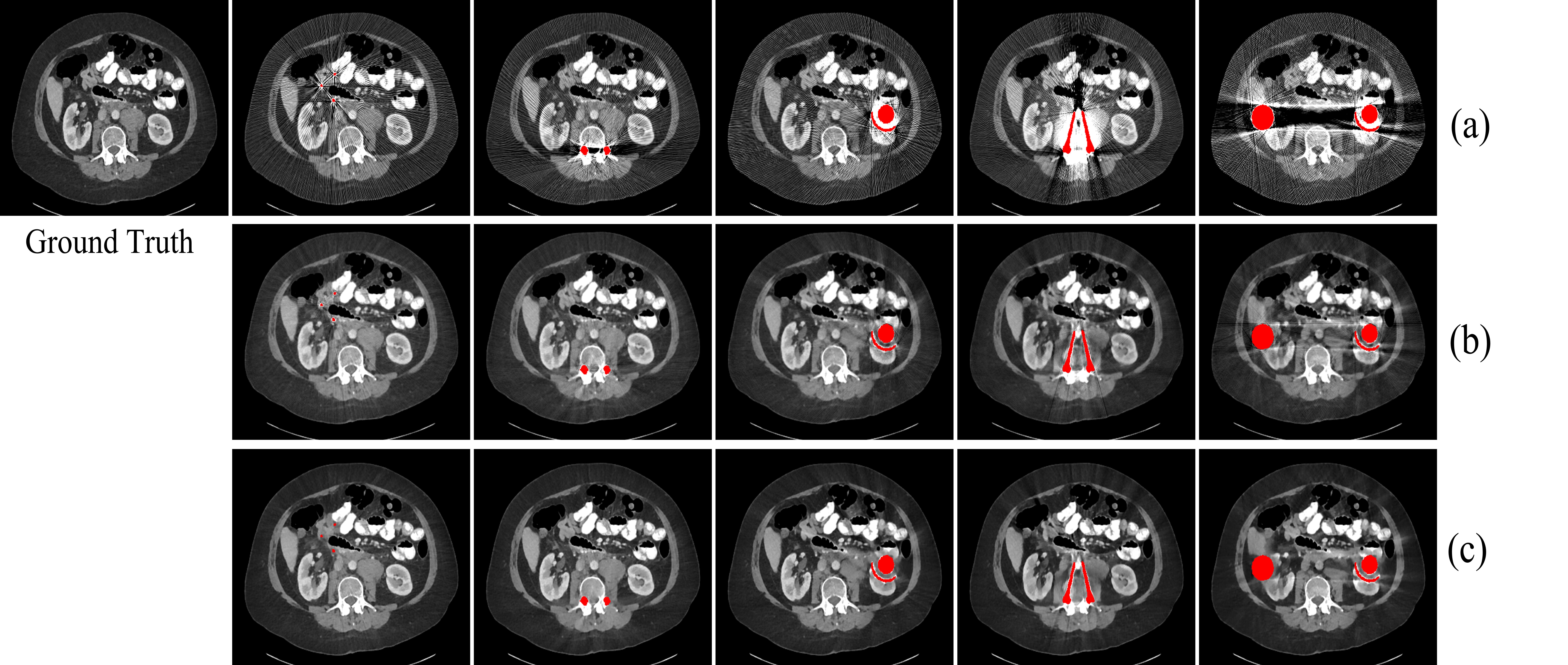}
	\end{center}
		\vspace{-2.5mm}
		\caption{\textcolor{black}{(a) Ground truth and metal-corrupted CT images with different metallic implants selected from synthesized DeepLesion; (b) The artifact-reduced results recovered by ``$\text{InDuDoNet+}_{{fbp}}$''; (c) The images predicted by InDuDoNet+. The red pixels stand for metals.}}
	\label{fig:xfbp}
\end{figure*}

\vspace{3mm}
\noindent{\textcolor{black}{\textbf{Number of Stages $N$.} Table~\ref{tab:n} lists the performance of our framework under different numbers of stages $N$ on the synthesized DeepLesion.
The $N=0$ entry means that the initialization $X_0$ is directly regarded as the reconstruction result, which is the same as the operation in InDuDoNet. Taking $N=0$ as the baseline, we can find that with only one stage ($N=1$), the MAR performance yielded by our proposed InDuDoNet+ is already evidently improved, which validates the effectiveness of the proposed mutual learning mechanism between $\widetilde{S}$-net and $X$-net. When $N=12$, the improvement magnitude of PSNR/SSIM is small. Hence, for the trade-off between MAR performance and the number of network parameters, we choose $N=10$ in all our experiments.}}

\vspace{3mm}
{\noindent{\textcolor{black}{\textcolor{black}{\textbf{Clinical Segmentation Mask.}}
 To  evaluate the impact  of the metal segmentation mask, Fig.~\ref{fig:mask} displays the artifact-reduced CT images on the clinical SpineWeb dataset where the clinical metal mask is segmented based on different thresholds, including 500~HU, 1000~HU, 2500~HU, and 4000~HU. As seen, a small thresholding (\emph{e.g.}, 500~HU) makes normal tissues be regarded as metal regions and the reconstructed images lose massive details; a large thresholding (\emph{e.g.}, 4000~HU) makes metallic implants be mistaken as body tissues and metal artifacts cannot be removed completely. Comparatively, the choice of the empirically-designed thresholding 2500~HU achieves the better image detail preservation and artifact reduction. As seen, it is very important to select the proper segmentation thresholding. In the future work, it is worthwhile to design an automatic metal localization algorithm for the better artifact-reduced CT image reconstruction.}}}

\vspace{3mm}
{\noindent{\textcolor{black}{\textcolor{black}{\textbf{Ways to Reconstruction.}} Table~\ref{tab:sx} reports the comparisons between $\text{InDuDoNet+}_{{fbp}}$ and $\text{InDuDoNet+}$, where $\text{InDuDoNet+}_{{fbp}}$ denotes the case that the filtered back projection of the reconstructed sinogram $S_{N}$ is taken as the final output image, and $\text{InDuDoNet+}$ we adopt in all comparison experiments denotes the case that the reconstructed image $X_{N}$ estimated by $X$-net is taken as the final artifact-reduced output. As seen, there is an evident performance difference between the two outputs. Fig.~\ref{fig:xfbp} presents the visual comparison on several artifact-affected CT images with different sizes of metallic implants randomly selected from the synthesized DeepLesion dataset. It is clearly observed that compared to $\text{InDuDoNet+}_{{fbp}}$, $\text{InDuDoNet+}$ removes more artifacts and achieves better reconstruction results. The underlying reasons are: 1) As derived in Sec.~\ref{sec:details}, at the last iterative stage, the reconstructed CT image $X_{N}$ is derived by the computation process contained in $X$-net as: $X_{N} = \text{proxNet}_{\theta_{x}^{(N)}}\left(X_{N-1}- \eta_{2}\mP^{T}\left(\mP X_{N-1}-{S}_{N}\right)\right)$. As seen, compared to the filtered back-projection of $S_{N}$, the output $X_{N}$ has higher flexibility due to the powerful representation ability and the high non-linearity of the proximal network $\text{proxNet}_{\theta_{x}^{(N)}}\left(\cdot\right)$. Then, the final $X$-net output would have the potential to achieve better MAR performance; 2) At the discretized space, the forward-projection and back-projection are not completely reversible to each other due to various approximations and numerical errors. In our model, we only incorporate the forward projection; therefore, the model is optimized for the forward projection. Recovering $X$ as a back-projection of $S$ is not as accurate as using $X$ itself directly.}}}


\subsection{Model Verification}
\begin{figure*}[!t]
  \begin{center}
     \includegraphics[width=1\linewidth]{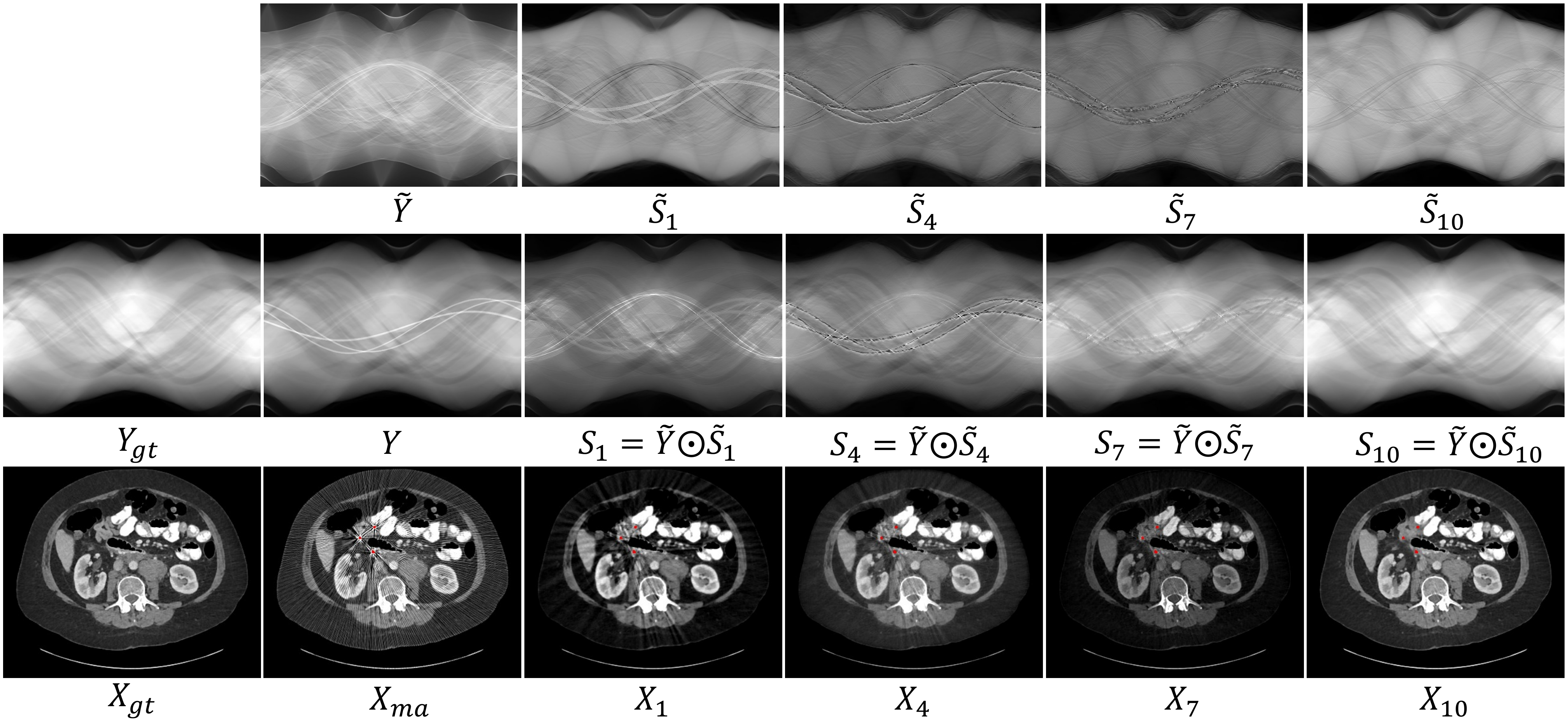}
  \end{center}
  \vspace{-6mm}
     \caption{The normalization coefficient $\widetilde{Y}$, normalized sinogram $\widetilde{S}_{n}$, sinogram $S_{n}$, and CT image $X_{n}$ restored by InDuDoNet+ at different stages where the number $N$ of the total iterative stages is 10. The red pixels stand for metallic implant.}
  \label{figmodelver}
    \vspace{1mm}
\end{figure*}

Here, we utilize InDuDoNet+ to execute a model verification experiment in order to present the working mechanism underlying the network modules ($\widetilde{S}$-net and $X$-net). 
Fig.~\ref{figmodelver} displays the reconstructed normalized sinogram $\widetilde{S}_{n}$, sinogram $S_{n}$, and CT image $X_{n}$ at different stages ($n=1,4,7,10$). It can be easily observed that with the increasing of $n$, the metal trace region in $\widetilde{S}_n$ is gradually flattened, which correspondingly ameliorates the sinogram $S_n$. Thus, the metal artifacts contained in the CT image $X_n$ are gradually removed. {\textcolor{black}{The results verify the design of our optimization-inspired iterative learning framework}}---the mutual promotion of $\widetilde{S}$-net and $X$-net enables the proposed InDuDoNet+ to achieve the MAR along the direction specified by Eq.~(\ref{o3}). {\textcolor{black}{Through this visualization result, the underlying rationality and insights of the proposed network can be intuitively understood by general users. As compared with most of heuristic network structures, our network has better transparency.}}

\subsection{Network Parameters and Inference Time}
For the compared MAR methods (\emph{i.e.,} DuDoNet, DSCMAR, DuDoNet++, InDuDoNet, and InDuDoNet+), Table~\ref{tabtime} lists the number of network parameters and the average inference time computed on 2,000 images with size $416\times416$ pixels on an NVIDIA Tesla V100-SMX2 GPU. As compared with other SOTA methods, the previously designed InDuDoNet has evidently fewer parameters, while the proposed InDuDoNet+ further reduces the network capacity. For inference time, InDuDoNet+ is comparable to DSCMAR, while faster than others. It is clear that due to the simple design of Prior-net, InDuDoNet+ performs better than previous InDuDoNet on  computational efficiency.



\begin{table}[!t]
\centering
\caption{Numbers of network parameters and average inference time (seconds) for different MAR methods.}
\setlength{\tabcolsep}{7.0pt}
\begin{tabular}{c|c|c|c|c|c}
\Xhline{0.6pt}
Methods    &DuDoNet   & DSCMAR &  DuDoNet++ & InDuDoNet&InDuDoNet+  \\
\hline
\# Network Parameters  &25,834,251 &25,834,251 & 25,983,627&5,174,936 & 1,782,007 \\
Inference Time (Seconds)   &0.4225  &0.3638 & 0.8062 & 0.5116 & 0.3782\\ 
\Xhline{0.6pt}
\end{tabular}
\label{tabtime}
\end{table}
\section{Conclusion and Future Work}\label{sec:conclu}
In this paper, for this metal artifact reduction (MAR) task, we have proposed a novel joint spatial and Radon domain reconstruction model and designed an optimization algorithm for solving it. By unfolding every iterative step into the corresponding network module, {\textcolor{black}{we constructed an optimization-inspired network architecture, namely InDuDoNet+.}} Besides, we analyzed the characteristics of metal-corrupted CT images and embedded such prior observations into our framework, which has clear interpretability and fine generalization ability. Comprehensive experiments conducted on synthesized and clinical data have substantiated the effectiveness of our dual-domain MAR approaches beyond current SOTA deep MAR networks.

As stated in Sec.~\ref{thre}, following the current SOTA methods, we adopted the simple thresholding to coarsely segment the metallic implants for clinical data, which is not very accurate and lacks flexibility. 
{\textcolor{black}{An unsatisfactory thresholding possibly makes tissues be wrongly regarded as metals and most MAR methods as well as our InDuDoNet+ would fail to recover image details. For performance improvement, in the future work, we will try to design an automatic metal localization algorithm and incorporate it into the proposed dual domain network framework.}} Besides, how to finely apply such model-driven dual domain framework in a semi-/un-supervised manner for better generalization performance would be an interesting research direction worthy of further exploration.

\vspace{2mm}
\noindent\textbf{Acknowledgements.} 
This research is supported by the China NSFC projects under contracts U21A6005, 61721002, U1811461, 12101061, the Major Key Project of PCL (PCL2021A12), and the Macao Science and Technology Development Fund under Grant 061/2020/A2.

\bibliography{egbib}

\begin{thebibliography}{45}
\expandafter\ifx\csname natexlab\endcsname\relax\def\natexlab#1{#1}\fi
\providecommand{\url}[1]{\texttt{#1}}
\providecommand{\href}[2]{#2}
\providecommand{\path}[1]{#1}
\providecommand{\DOIprefix}{doi:}
\providecommand{\ArXivprefix}{arXiv:}
\providecommand{\URLprefix}{URL: }
\providecommand{\Pubmedprefix}{pmid:}
\providecommand{\doi}[1]{\href{http://dx.doi.org/#1}{\path{#1}}}
\providecommand{\Pubmed}[1]{\href{pmid:#1}{\path{#1}}}
\providecommand{\bibinfo}[2]{#2}
\ifx\xfnm\relax \def\xfnm[#1]{\unskip,\space#1}\fi
\bibitem[{Beck and Teboulle(2009)}]{beck2009fast}
\bibinfo{author}{Beck, A.}, \bibinfo{author}{Teboulle, M.},
  \bibinfo{year}{2009}.
\newblock \bibinfo{title}{A fast iterative shrinkage-thresholding algorithm for
  linear inverse problems}.
\newblock \bibinfo{journal}{SIAM Journal on Imaging Sciences}
  \bibinfo{volume}{2}, \bibinfo{pages}{183--202}.
\bibitem[{Chang et~al.(2018)Chang, Ye, Srivastava, Thibault, Sauer and
  Bouman}]{chang2018prior}
\bibinfo{author}{Chang, Z.}, \bibinfo{author}{Ye, D.H.},
  \bibinfo{author}{Srivastava, S.}, \bibinfo{author}{Thibault, J.B.},
  \bibinfo{author}{Sauer, K.}, \bibinfo{author}{Bouman, C.},
  \bibinfo{year}{2018}.
\newblock \bibinfo{title}{Prior-guided metal artifact reduction for iterative
  {X}-ray computed tomography}.
\newblock \bibinfo{journal}{IEEE {T}ransactions on {M}edical {I}maging}
  \bibinfo{volume}{38}, \bibinfo{pages}{1532--1542}.
\bibitem[{De~Man et~al.(1999)De~Man, Nuyts, Dupont, Marchal and
  Suetens}]{de1999metal}
\bibinfo{author}{De~Man, B.}, \bibinfo{author}{Nuyts, J.},
  \bibinfo{author}{Dupont, P.}, \bibinfo{author}{Marchal, G.},
  \bibinfo{author}{Suetens, P.}, \bibinfo{year}{1999}.
\newblock \bibinfo{title}{Metal streak artifacts in {X}-ray computed
  tomography: A simulation study}.
\newblock \bibinfo{journal}{IEEE {T}ransactions on {N}uclear {S}cience}
  \bibinfo{volume}{46}, \bibinfo{pages}{691--696}.
\bibitem[{Donoho(1995)}]{donoho1995noising}
\bibinfo{author}{Donoho, D.L.}, \bibinfo{year}{1995}.
\newblock \bibinfo{title}{De-noising by soft-thresholding}.
\newblock \bibinfo{journal}{IEEE {T}ransactions on {I}nformation {T}heory}
  \bibinfo{volume}{41}, \bibinfo{pages}{613--627}.
\bibitem[{Fu et~al.(2022)Fu, Wang, Xie, Zhao, Meng and Xu}]{fu2022kxnet}
\bibinfo{author}{Fu, J.}, \bibinfo{author}{Wang, H.}, \bibinfo{author}{Xie,
  Q.}, \bibinfo{author}{Zhao, Q.}, \bibinfo{author}{Meng, D.},
  \bibinfo{author}{Xu, Z.}, \bibinfo{year}{2022}.
\newblock \bibinfo{title}{{KXN}et: {A} model-driven deep neural network for
  blind super-resolution}.
\newblock \bibinfo{journal}{arXiv preprint arXiv:2209.10305} .
\bibitem[{Ghani and Karl(2019)}]{ghani2019fast}
\bibinfo{author}{Ghani, M.U.}, \bibinfo{author}{Karl, W.C.},
  \bibinfo{year}{2019}.
\newblock \bibinfo{title}{Fast enhanced {C}{T} metal artifact reduction using
  data domain deep learning}.
\newblock \bibinfo{journal}{IEEE {T}ransactions on {C}omputational {I}maging}
  \bibinfo{volume}{6}, \bibinfo{pages}{181--193}.
\bibitem[{Gjesteby et~al.(2017)Gjesteby, Yang, Xi, Zhou, Zhang and
  Wang}]{gjesteby2017deep}
\bibinfo{author}{Gjesteby, L.}, \bibinfo{author}{Yang, Q.},
  \bibinfo{author}{Xi, Y.}, \bibinfo{author}{Zhou, Y.}, \bibinfo{author}{Zhang,
  J.}, \bibinfo{author}{Wang, G.}, \bibinfo{year}{2017}.
\newblock \bibinfo{title}{Deep learning methods to guide {C}{T} image
  reconstruction and reduce metal artifacts}, in: \bibinfo{booktitle}{Medical
  Imaging: Physics of Medical Imaging}, \bibinfo{organization}{International
  Society for Optics and Photonics}. p. \bibinfo{pages}{101322W}.
\bibitem[{He et~al.(2016)He, Zhang, Ren and Sun}]{he2016deep}
\bibinfo{author}{He, K.}, \bibinfo{author}{Zhang, X.}, \bibinfo{author}{Ren,
  S.}, \bibinfo{author}{Sun, J.}, \bibinfo{year}{2016}.
\newblock \bibinfo{title}{Deep residual learning for image recognition}, in:
  \bibinfo{booktitle}{Proceedings of the IEEE Conference on Computer Vision and
  Pattern Recognition}, pp. \bibinfo{pages}{770--778}.
\bibitem[{Huang et~al.(2018)Huang, Wang, Tang, Zhong and
  Zhang}]{huang2018metal}
\bibinfo{author}{Huang, X.}, \bibinfo{author}{Wang, J.}, \bibinfo{author}{Tang,
  F.}, \bibinfo{author}{Zhong, T.}, \bibinfo{author}{Zhang, Y.},
  \bibinfo{year}{2018}.
\newblock \bibinfo{title}{Metal artifact reduction on cervical {C}{T} images by
  deep residual learning}.
\newblock \bibinfo{journal}{Biomedical {E}ngineering {O}nline}
  \bibinfo{volume}{17}, \bibinfo{pages}{1--15}.
\bibitem[{Jin et~al.(2015)Jin, Bouman and Sauer}]{jin2015model}
\bibinfo{author}{Jin, P.}, \bibinfo{author}{Bouman, C.A.},
  \bibinfo{author}{Sauer, K.D.}, \bibinfo{year}{2015}.
\newblock \bibinfo{title}{A model-based image reconstruction algorithm with
  simultaneous beam hardening correction for {X}-ray {C}{T}}.
\newblock \bibinfo{journal}{IEEE Transactions on Computational Imaging}
  \bibinfo{volume}{1}, \bibinfo{pages}{200--216}.
\bibitem[{Kalender et~al.(1987)Kalender, Hebel and
  Ebersberger}]{kalender1987reduction}
\bibinfo{author}{Kalender, W.A.}, \bibinfo{author}{Hebel, R.},
  \bibinfo{author}{Ebersberger, J.}, \bibinfo{year}{1987}.
\newblock \bibinfo{title}{Reduction of {C}{T} artifacts caused by metallic
  implants}.
\newblock \bibinfo{journal}{Radiology} \bibinfo{volume}{164},
  \bibinfo{pages}{576--577}.
\bibitem[{Karimi et~al.(2015)Karimi, Martz and Cosman}]{karimi2015metal}
\bibinfo{author}{Karimi, S.}, \bibinfo{author}{Martz, H.},
  \bibinfo{author}{Cosman, P.}, \bibinfo{year}{2015}.
\newblock \bibinfo{title}{Metal artifact reduction for {C}{T}-based luggage
  screening}.
\newblock \bibinfo{journal}{Journal of X-ray Science and Technology}
  \bibinfo{volume}{23}, \bibinfo{pages}{435--451}.
\bibitem[{Lemmens et~al.(2008)Lemmens, Faul and Nuyts}]{lemmens2008suppression}
\bibinfo{author}{Lemmens, C.}, \bibinfo{author}{Faul, D.},
  \bibinfo{author}{Nuyts, J.}, \bibinfo{year}{2008}.
\newblock \bibinfo{title}{Suppression of metal artifacts in {C}{T} using a
  reconstruction procedure that combines {M}{A}{P} and projection completion}.
\newblock \bibinfo{journal}{IEEE {T}ansactions on {M}edical {I}maging}
  \bibinfo{volume}{28}, \bibinfo{pages}{250--260}.
\bibitem[{Liao et~al.(2019a)Liao, Lin, Huo, Vogelsang, Sehnert, Zhou and
  Luo}]{liao2019generative}
\bibinfo{author}{Liao, H.}, \bibinfo{author}{Lin, W.A.}, \bibinfo{author}{Huo,
  Z.}, \bibinfo{author}{Vogelsang, L.}, \bibinfo{author}{Sehnert, W.J.},
  \bibinfo{author}{Zhou, S.K.}, \bibinfo{author}{Luo, J.},
  \bibinfo{year}{2019}a.
\newblock \bibinfo{title}{Generative mask pyramid network for
  {C}{T}/{C}{B}{C}{T} metal artifact reduction with joint projection-sinogram
  correction}, in: \bibinfo{booktitle}{International Conference on Medical
  Image Computing and Computer Assisted Intervention}, pp.
  \bibinfo{pages}{77--85}.
\bibitem[{Liao et~al.(2019b)Liao, Lin, Zhou and Luo}]{liao2019adn}
\bibinfo{author}{Liao, H.}, \bibinfo{author}{Lin, W.A.}, \bibinfo{author}{Zhou,
  S.K.}, \bibinfo{author}{Luo, J.}, \bibinfo{year}{2019}b.
\newblock \bibinfo{title}{{A}{D}{N}: Artifact disentanglement network for
  unsupervised metal artifact reduction}.
\newblock \bibinfo{journal}{IEEE Transactions on Medical Imaging}
  \bibinfo{volume}{39}, \bibinfo{pages}{634--643}.
\bibitem[{Lin et~al.(2019)Lin, Liao, Peng, Sun, Zhang, Luo, Chellappa and
  Zhou}]{lin2019dudonet}
\bibinfo{author}{Lin, W.A.}, \bibinfo{author}{Liao, H.}, \bibinfo{author}{Peng,
  C.}, \bibinfo{author}{Sun, X.}, \bibinfo{author}{Zhang, J.},
  \bibinfo{author}{Luo, J.}, \bibinfo{author}{Chellappa, R.},
  \bibinfo{author}{Zhou, S.K.}, \bibinfo{year}{2019}.
\newblock \bibinfo{title}{Du{D}o{N}et: Dual domain network for {C}{T} metal
  artifact reduction}, in: \bibinfo{booktitle}{Proceedings of the IEEE/CVF
  Conference on Computer Vision and Pattern Recognition}, pp.
  \bibinfo{pages}{10512--10521}.
\bibitem[{Liu et~al.(2022)Liu, Xie, Zhao, Wang and Meng}]{liu2022low}
\bibinfo{author}{Liu, X.}, \bibinfo{author}{Xie, Q.}, \bibinfo{author}{Zhao,
  Q.}, \bibinfo{author}{Wang, H.}, \bibinfo{author}{Meng, D.},
  \bibinfo{year}{2022}.
\newblock \bibinfo{title}{{Low-light image enhancement by {R}etinex based
  algorithm unrolling and adjustment}}.
\newblock \bibinfo{journal}{arXiv preprint arXiv:2202.05972} .
\bibitem[{Lyu et~al.(2020)Lyu, Lin, Liao, Lu and Zhou}]{lyu2020dudonet++}
\bibinfo{author}{Lyu, Y.}, \bibinfo{author}{Lin, W.A.}, \bibinfo{author}{Liao,
  H.}, \bibinfo{author}{Lu, J.}, \bibinfo{author}{Zhou, S.K.},
  \bibinfo{year}{2020}.
\newblock \bibinfo{title}{Encoding metal mask projection for metal artifact
  reduction in computed tomography}, in: \bibinfo{booktitle}{International
  {C}onference on {M}edical {I}mage {C}omputing and {C}omputer-{A}ssisted
  {I}ntervention}, pp. \bibinfo{pages}{147--157}.
\bibitem[{Mehranian et~al.(2013)Mehranian, Ay, Rahmim and
  Zaidi}]{mehranian2013x}
\bibinfo{author}{Mehranian, A.}, \bibinfo{author}{Ay, M.R.},
  \bibinfo{author}{Rahmim, A.}, \bibinfo{author}{Zaidi, H.},
  \bibinfo{year}{2013}.
\newblock \bibinfo{title}{{X}-ray {C}{T} metal artifact reduction using wavelet
  domain $ l_0$ sparse regularization}.
\newblock \bibinfo{journal}{IEEE {T}ransactions on {M}edical {I}maging}
  \bibinfo{volume}{32}, \bibinfo{pages}{1707--1722}.
\bibitem[{Meyer et~al.(2010)Meyer, Raupach, Lell, Schmidt and
  Kachelrie{\ss}}]{meyer2010normalized}
\bibinfo{author}{Meyer, E.}, \bibinfo{author}{Raupach, R.},
  \bibinfo{author}{Lell, M.}, \bibinfo{author}{Schmidt, B.},
  \bibinfo{author}{Kachelrie{\ss}, M.}, \bibinfo{year}{2010}.
\newblock \bibinfo{title}{Normalized metal artifact reduction ({N}{M}{A}{R}) in
  computed tomography}.
\newblock \bibinfo{journal}{Medical Physics} \bibinfo{volume}{37},
  \bibinfo{pages}{5482--5493}.
\bibitem[{Park et~al.(2018)Park, Lee, Kim, Seo and Chung}]{park2018ct}
\bibinfo{author}{Park, H.S.}, \bibinfo{author}{Lee, S.M.},
  \bibinfo{author}{Kim, H.P.}, \bibinfo{author}{Seo, J.K.},
  \bibinfo{author}{Chung, Y.E.}, \bibinfo{year}{2018}.
\newblock \bibinfo{title}{{C}{T} sinogram-consistency learning for
  metal-induced beam hardening correction}.
\newblock \bibinfo{journal}{Medical Physics} \bibinfo{volume}{45},
  \bibinfo{pages}{5376--5384}.
\bibitem[{Paszke et~al.(2019)Paszke, Gross, Massa, Lerer, Bradbury, Chanan,
  Killeen, Lin, Gimelshein, Antiga et~al.}]{paszke2017automatic}
\bibinfo{author}{Paszke, A.}, \bibinfo{author}{Gross, S.},
  \bibinfo{author}{Massa, F.}, \bibinfo{author}{Lerer, A.},
  \bibinfo{author}{Bradbury, J.}, \bibinfo{author}{Chanan, G.},
  \bibinfo{author}{Killeen, T.}, \bibinfo{author}{Lin, Z.},
  \bibinfo{author}{Gimelshein, N.}, \bibinfo{author}{Antiga, L.}, et~al.,
  \bibinfo{year}{2019}.
\newblock \bibinfo{title}{Pytorch: {A}n imperative style, high-performance deep
  learning library}.
\newblock \bibinfo{journal}{Advances in Neural Information Processing Systems}
  \bibinfo{volume}{32}, \bibinfo{pages}{8026--8037}.
\bibitem[{Ronneberger et~al.(2015)Ronneberger, Fischer and
  Brox}]{ronneberger2015u}
\bibinfo{author}{Ronneberger, O.}, \bibinfo{author}{Fischer, P.},
  \bibinfo{author}{Brox, T.}, \bibinfo{year}{2015}.
\newblock \bibinfo{title}{U-{N}et: Convolutional networks for biomedical image
  segmentation}, in: \bibinfo{booktitle}{International Conference on Medical
  Image Computing and Computer Assisted Intervention}, pp.
  \bibinfo{pages}{234--241}.
\bibitem[{Schiffer and Bredies(2014)}]{schiffer2014sinogram}
\bibinfo{author}{Schiffer, C.}, \bibinfo{author}{Bredies, K.},
  \bibinfo{year}{2014}.
\newblock \bibinfo{title}{Sinogram constrained {T}{V}-minimization for metal
  artifact reduction in {C}{T}}.
\newblock \bibinfo{journal}{arXiv preprint arXiv:1404.6691} .
\bibitem[{Soltanian-Zadeh et~al.(1996)Soltanian-Zadeh, Windham and
  Soltanianzadeh}]{soltanian1996ct}
\bibinfo{author}{Soltanian-Zadeh, H.}, \bibinfo{author}{Windham, J.P.},
  \bibinfo{author}{Soltanianzadeh, J.}, \bibinfo{year}{1996}.
\newblock \bibinfo{title}{{C}{T} artifact correction: an image-processing
  approach}, in: \bibinfo{booktitle}{Medical Imaging: Image Processing},
  \bibinfo{organization}{International Society for Optics and Photonics}. pp.
  \bibinfo{pages}{477--485}.
\bibitem[{Wang et~al.(2018a)Wang, Ye, Mueller and Fessler}]{wang2018image}
\bibinfo{author}{Wang, G.}, \bibinfo{author}{Ye, J.C.},
  \bibinfo{author}{Mueller, K.}, \bibinfo{author}{Fessler, J.A.},
  \bibinfo{year}{2018}a.
\newblock \bibinfo{title}{Image reconstruction is a new frontier of machine
  learning}.
\newblock \bibinfo{journal}{IEEE Transactions on Medical Imaging}
  \bibinfo{volume}{37}, \bibinfo{pages}{1289--1296}.
\bibitem[{Wang et~al.(2021a)Wang, Li, He, Ma, Meng and Zheng}]{wang2021dicdnet}
\bibinfo{author}{Wang, H.}, \bibinfo{author}{Li, Y.}, \bibinfo{author}{He, N.},
  \bibinfo{author}{Ma, K.}, \bibinfo{author}{Meng, D.}, \bibinfo{author}{Zheng,
  Y.}, \bibinfo{year}{2021}a.
\newblock \bibinfo{title}{{DICDN}et: Deep interpretable convolutional
  dictionary network for metal artifact reduction in {CT} images}.
\newblock \bibinfo{journal}{IEEE Transactions on Medical Imaging}
  \bibinfo{volume}{41}, \bibinfo{pages}{869--880}.
\bibitem[{Wang et~al.(2022a)Wang, Li, Meng and Zheng}]{wang2022adaptive}
\bibinfo{author}{Wang, H.}, \bibinfo{author}{Li, Y.}, \bibinfo{author}{Meng,
  D.}, \bibinfo{author}{Zheng, Y.}, \bibinfo{year}{2022}a.
\newblock \bibinfo{title}{Adaptive convolutional dictionary network for {CT}
  metal artifact reduction}.
\newblock \bibinfo{journal}{arXiv preprint arXiv:2205.07471} .
\bibitem[{Wang et~al.(2021b)Wang, Li, Zhang, Chen, Ma, Meng and
  Zheng}]{wang2021indudonet}
\bibinfo{author}{Wang, H.}, \bibinfo{author}{Li, Y.}, \bibinfo{author}{Zhang,
  H.}, \bibinfo{author}{Chen, J.}, \bibinfo{author}{Ma, K.},
  \bibinfo{author}{Meng, D.}, \bibinfo{author}{Zheng, Y.},
  \bibinfo{year}{2021}b.
\newblock \bibinfo{title}{{InDuDoNet}: An interpretable dual domain network for
  {CT} metal artifact reduction}, in: \bibinfo{booktitle}{International
  Conference on Medical Image Computing and Computer Assisted Intervention},
  \bibinfo{organization}{Springer}. pp. \bibinfo{pages}{107--118}.
\bibitem[{Wang et~al.(2022b)Wang, Xie, Li, Huang, Meng and
  Zheng}]{wang2022orientation}
\bibinfo{author}{Wang, H.}, \bibinfo{author}{Xie, Q.}, \bibinfo{author}{Li,
  Y.}, \bibinfo{author}{Huang, Y.}, \bibinfo{author}{Meng, D.},
  \bibinfo{author}{Zheng, Y.}, \bibinfo{year}{2022}b.
\newblock \bibinfo{title}{Orientation-shared convolution representation for
  {CT} metal artifact learning}, in: \bibinfo{booktitle}{International
  Conference on Medical Image Computing and Computer-Assisted Intervention},
  pp. \bibinfo{pages}{665--675}.
\bibitem[{Wang et~al.(2021c)Wang, Xie, Zhao, Liang and Meng}]{wang2021rcdnet}
\bibinfo{author}{Wang, H.}, \bibinfo{author}{Xie, Q.}, \bibinfo{author}{Zhao,
  Q.}, \bibinfo{author}{Liang, Y.}, \bibinfo{author}{Meng, D.},
  \bibinfo{year}{2021}c.
\newblock \bibinfo{title}{{RCDNet}: An interpretable rain convolutional
  dictionary network for single image deraining}.
\newblock \bibinfo{journal}{arXiv preprint arXiv:2107.06808} .
\bibitem[{Wang et~al.(2020)Wang, Xie, Zhao and Meng}]{wang2020model}
\bibinfo{author}{Wang, H.}, \bibinfo{author}{Xie, Q.}, \bibinfo{author}{Zhao,
  Q.}, \bibinfo{author}{Meng, D.}, \bibinfo{year}{2020}.
\newblock \bibinfo{title}{A model-driven deep neural network for single image
  rain removal}, in: \bibinfo{booktitle}{Proceedings of the IEEE/CVF Conference
  on Computer Vision and Pattern Recognition}, pp. \bibinfo{pages}{3103--3112}.
\bibitem[{Wang et~al.(2013)Wang, Wang, Chen, Wu, Coatrieux and
  Luo}]{wang2013metal}
\bibinfo{author}{Wang, J.}, \bibinfo{author}{Wang, S.}, \bibinfo{author}{Chen,
  Y.}, \bibinfo{author}{Wu, J.}, \bibinfo{author}{Coatrieux, J.L.},
  \bibinfo{author}{Luo, L.}, \bibinfo{year}{2013}.
\newblock \bibinfo{title}{Metal artifact reduction in {C}{T} using fusion based
  prior image}.
\newblock \bibinfo{journal}{Medical Physics} \bibinfo{volume}{40},
  \bibinfo{pages}{081903}.
\bibitem[{Wang et~al.(2018b)Wang, Zhao, Noble and Dawant}]{wang2018conditional}
\bibinfo{author}{Wang, J.}, \bibinfo{author}{Zhao, Y.}, \bibinfo{author}{Noble,
  J.H.}, \bibinfo{author}{Dawant, B.M.}, \bibinfo{year}{2018}b.
\newblock \bibinfo{title}{Conditional generative adversarial networks for metal
  artifact reduction in {C}{T} images of the ear}, in:
  \bibinfo{booktitle}{International Conference on Medical Image Computing and
  Computer Assisted Intervention}, pp. \bibinfo{pages}{3--11}.
\bibitem[{Xie et~al.(2019)Xie, Zhou, Zhao, Meng, Zuo and Xu}]{xie2020mhf}
\bibinfo{author}{Xie, Q.}, \bibinfo{author}{Zhou, M.}, \bibinfo{author}{Zhao,
  Q.}, \bibinfo{author}{Meng, D.}, \bibinfo{author}{Zuo, W.},
  \bibinfo{author}{Xu, Z.}, \bibinfo{year}{2019}.
\newblock \bibinfo{title}{Multispectral and hyperspectral image fusion by
  {MS/HS} fusion net}, in: \bibinfo{booktitle}{Proceedings of the IEEE/CVF
  Conference on Computer Vision and Pattern Recognition}, pp.
  \bibinfo{pages}{1585--1594}.
\bibitem[{Yan et~al.(2018)Yan, Wang, Lu, Zhang, Harrison, Bagheri and
  Summers}]{yan2018deep}
\bibinfo{author}{Yan, K.}, \bibinfo{author}{Wang, X.}, \bibinfo{author}{Lu,
  L.}, \bibinfo{author}{Zhang, L.}, \bibinfo{author}{Harrison, A.P.},
  \bibinfo{author}{Bagheri, M.}, \bibinfo{author}{Summers, R.M.},
  \bibinfo{year}{2018}.
\newblock \bibinfo{title}{Deep lesion graphs in the wild: Relationship learning
  and organization of significant radiology image findings in a diverse
  large-scale lesion database}, in: \bibinfo{booktitle}{Proceedings of the IEEE
  Conference on Computer Vision and Pattern Recognition}, pp.
  \bibinfo{pages}{9261--9270}.
\bibitem[{Yang and Sun(2018)}]{yang2018proximal}
\bibinfo{author}{Yang, D.}, \bibinfo{author}{Sun, J.}, \bibinfo{year}{2018}.
\newblock \bibinfo{title}{Proximal {D}ehaze-{N}et: A prior learning-based deep
  network for single image dehazing}, in: \bibinfo{booktitle}{Proceedings of
  the European Conference on Computer Vision}, pp. \bibinfo{pages}{702--717}.
\bibitem[{Yang et~al.(2017)Yang, Sun, Li and Xu}]{yang2017admm}
\bibinfo{author}{Yang, Y.}, \bibinfo{author}{Sun, J.}, \bibinfo{author}{Li,
  H.}, \bibinfo{author}{Xu, Z.}, \bibinfo{year}{2017}.
\newblock \bibinfo{title}{A{D}{M}{M}-{N}et: A deep learning approach for
  compressive sensing {M}{R}{I}}.
\newblock \bibinfo{journal}{arXiv preprint arXiv:1705.06869} .
\bibitem[{Yu et~al.(2020)Yu, Zhang, Li and Xing}]{yu2020deep}
\bibinfo{author}{Yu, L.}, \bibinfo{author}{Zhang, Z.}, \bibinfo{author}{Li,
  X.}, \bibinfo{author}{Xing, L.}, \bibinfo{year}{2020}.
\newblock \bibinfo{title}{Deep sinogram completion with image prior for metal
  artifact reduction in {C}{T} images}.
\newblock \bibinfo{journal}{IEEE Transactions on Medical Imaging}
  \bibinfo{volume}{40}, \bibinfo{pages}{228--238}.
\bibitem[{Zhang et~al.(2018)Zhang, Dong and Liu}]{zhang2018reweighted}
\bibinfo{author}{Zhang, H.}, \bibinfo{author}{Dong, B.}, \bibinfo{author}{Liu,
  B.}, \bibinfo{year}{2018}.
\newblock \bibinfo{title}{A reweighted joint spatial-{R}adon domain {C}{T}
  image reconstruction model for metal artifact reduction}.
\newblock \bibinfo{journal}{{S}{I}{A}{M} {J}ournal on {I}maging {S}ciences}
  \bibinfo{volume}{11}, \bibinfo{pages}{707--733}.
\bibitem[{Zhang et~al.(2016)Zhang, Wang, Li, Cai, Hu and
  Yan}]{zhang2016iterative}
\bibinfo{author}{Zhang, H.}, \bibinfo{author}{Wang, L.}, \bibinfo{author}{Li,
  L.}, \bibinfo{author}{Cai, A.}, \bibinfo{author}{Hu, G.},
  \bibinfo{author}{Yan, B.}, \bibinfo{year}{2016}.
\newblock \bibinfo{title}{Iterative metal artifact reduction for {X}-ray
  computed tomography using unmatched projector/backprojector pairs}.
\newblock \bibinfo{journal}{Medical Physics} \bibinfo{volume}{43},
  \bibinfo{pages}{3019--3033}.
\bibitem[{Zhang and Ghanem(2018)}]{zhang2018ista}
\bibinfo{author}{Zhang, J.}, \bibinfo{author}{Ghanem, B.},
  \bibinfo{year}{2018}.
\newblock \bibinfo{title}{\textcolor{black}{{ISTA-N}et: Interpretable
  optimization-inspired deep network for image compressive sensing}}, in:
  \bibinfo{booktitle}{Proceedings of the IEEE Conference on Computer Vision and
  Pattern Recognition}, pp. \bibinfo{pages}{1828--1837}.
\bibitem[{Zhang et~al.(2020)Zhang, Gool and Timofte}]{zhang2020deep}
\bibinfo{author}{Zhang, K.}, \bibinfo{author}{Gool, L.V.},
  \bibinfo{author}{Timofte, R.}, \bibinfo{year}{2020}.
\newblock \bibinfo{title}{\textcolor{black}{Deep unfolding network for image
  super-resolution}}, in: \bibinfo{booktitle}{Proceedings of the IEEE/CVF
  Conference on Computer Vision and Pattern Recognition}, pp.
  \bibinfo{pages}{3217--3226}.
\bibitem[{Zhang and Yu(2018)}]{zhang2018convolutional}
\bibinfo{author}{Zhang, Y.}, \bibinfo{author}{Yu, H.}, \bibinfo{year}{2018}.
\newblock \bibinfo{title}{Convolutional neural network based metal artifact
  reduction in {X}-ray computed tomography}.
\newblock \bibinfo{journal}{IEEE Transactions on Medical Imaging}
  \bibinfo{volume}{37}, \bibinfo{pages}{1370--1381}.
\bibitem[{Zhou et~al.(2022)Zhou, Chen, Zhou, Duncan and Liu}]{zhou2022dudodr}
\bibinfo{author}{Zhou, B.}, \bibinfo{author}{Chen, X.}, \bibinfo{author}{Zhou,
  S.K.}, \bibinfo{author}{Duncan, J.S.}, \bibinfo{author}{Liu, C.},
  \bibinfo{year}{2022}.
\newblock \bibinfo{title}{\textcolor{black}{{DuDoDR-N}et: Dual-domain data
  consistent recurrent network for simultaneous sparse view and metal artifact
  reduction in computed tomography}}.
\newblock \bibinfo{journal}{Medical Image Analysis} \bibinfo{volume}{75},
  \bibinfo{pages}{102289}.

\end{thebibliography}

\end{document}